\documentstyle[epsfig,times]{ecco94}
\begin{document}


\def\trueital{
    \everymath={\it}
    \everydisplay={\it}
    \mathcode"2C="702C\mathcode"2E="702E
    \mathcode"3A="703A\mathcode"3B="703B
    \mathcode"30="0030 \mathcode"31="0031 \mathcode"32="0032
    \mathcode"33="0033 \mathcode"34="0034 \mathcode"35="0035
    \mathcode"36="0036 \mathcode"37="0037 \mathcode"38="0038
    \mathcode"39="0039
    \mathcode"41="7041\mathcode"42="7042\mathcode"43="7043
    \mathcode"44="7044\mathcode"45="7045\mathcode"46="7046
    \mathcode"47="7047\mathcode"48="7048\mathcode"49="7049
    \mathcode"4A="704A\mathcode"4B="704B\mathcode"4C="704C
    \mathcode"4D="704D\mathcode"4E="704E\mathcode"4F="704F
    \mathcode"50="7050\mathcode"51="7051\mathcode"52="7052
    \mathcode"53="7053\mathcode"54="7054\mathcode"55="7055
    \mathcode"56="7056\mathcode"57="7057\mathcode"58="7058
    \mathcode"59="7059\mathcode"5A="705A
    \mathcode"61="7061
    \mathcode"62="7062\mathcode"63="7063\mathcode"64="7064
    \mathcode"65="7065\mathcode"66="7066\mathcode"67="7067
    \mathcode"68="7068\mathcode"69="7069\mathcode"6A="706A
    \mathcode"6B="706B\mathcode"6C="706C\mathcode"6D="706D
    \mathcode"6E="706E\mathcode"6F="706F\mathcode"70="7070
    \mathcode"71="7071\mathcode"72="7072\mathcode"73="7073
    \mathcode"74="7074\mathcode"75="7075\mathcode"76="7076
    \mathcode"77="7077\mathcode"78="7078\mathcode"79="7079
    \mathcode"7A="707A
    \def\Gamma{\mathchar"0000}
    \def\Delta{\mathchar"0001}
    \def\Theta{\mathchar"0002}
    \def\Lambda{\mathchar"0003}
    \def\Xi{\mathchar"0004}
    \def\Pi{\mathchar"0005}
    \def\Sigma{\mathchar"0006}
    \def\Upsilon{\mathchar"0007}
    \def\Phi{\mathchar"0008}
    \def\Psi{\mathchar"0009}
    \def\Omega{\mathchar"000A}
}
\trueital

\input psfig
\pagestyle{empty}

\hyphenation{non-uni-que-ly}
\hyphenation{me-thod}
\hyphenation{Uly-sses}
\hyphenation{boun-da-ry}
\hyphenation{sche-mes}
\hyphenation{ma-the-ma-ti-cally}
\def\sgn{\mathop{\rm sgn}\nolimits}
\def\div{\mathop{\rm div}\nolimits}
\def\rot{\mathop{\rm rot}\nolimits}
\def\AU{{\rm AU}}
\def\const{\mathop{\rm const}\nolimits}
\def\minmod{\mathop{\rm min\,mod}\nolimits}
\def\tfrac#1#2{{\textstyle\frac{#1}{#2}}}
\def\fd#1#2{\frac{d#1}{d#2}}
\def\pd#1#2{\frac{\partial#1}{\partial#2}}
\def\brk#1{\ifx=#1=\hyphenation{}{}{}\fi}
\let\ds=\displaystyle
\def\diag{\mathop{\rm diag}\nolimits}
\def\det{\mathop{\rm det}\nolimits}
\def\sskip{\noalign{\kern 2pt}}
\mathchardef\0="7630

\title{Application of numerical methods to modeling \\
the stellar wind and interstellar medium interaction}

\author{
{\bf N. Pogorelov}
\institute{Department of  Earth \& Planetary Sciences, Kobe University, 1-1
Rokkodai-cho, Nada-ku, Kobe 657, JAPAN and
Institute for Problems in Mechanics, 101\ Vernadskii Ave.,
Moscow 117526, RUSSIA~(permanent address)}
\and
{\bf T. Matsuda}
\institute{Department of  Earth \& Planetary Sciences, Kobe University, 1-1
Rokkodai-cho, Nada-ku, Kobe 657, JAPAN}
}


\maketitle
\bibliographystyle{ecco94}

\begin{abstract}
Interaction between the stellar wind, including the solar wind, and the
interstellar medium has long been the subject of investigation by both
astrophysicists and fluid dynamicists. This is, first, due to the
possibility of comparison of physical models for such interaction with
the measurements performed by Voyager, Pioneer, and Ulysses spacecrafts.
On the other hand, a complicated structure of the flow containing several
discontinuities makes it a challenging problem for the application of
modern numerical methods both in gasdynamic and magnetogasdynamic (MHD)
cases.

In the solar wind case, the problem becomes even more
complicated, since the charge-exchange processes between ions and
neutral particles must be taken into account. The continuum equations
are not applicable to the description of the neutral particle motion,
for their mean free path is much larger than the characteristic length
scale of the problem. In this case, either approximate coupling models
or direct Monte-Carlo simulation are required. The spatial nonuniformity
of the solar wind and its perturbations and periodicity make the problem
three-dimensional and nonstationary. From a mechanical viewpoint the
problem represents the interaction of the uniform interstellar medium
and the spherically-symmetric (or asymmetric) solar wind flow.
We consider various approaches used by different authors to solve this
problem numerically.

The presence of the contact surface dividing the two flows
rises the question of its stability. We discuss the reasons of such
instabilities and parameters which influence it.

The presence of the interstellar magnetic field necessitates
solution of the MHD equations for proper analysis of the obtained data.
Although the system of governing equations remains hyperbolic in this
case, the multivariance of the exact solution to the MHD Riemann problem
makes inefficient its application for regular calculations. On the other
hand, the  solution to the linearized Riemann problem is
nonunique. We discuss the possible ways of applying the Roe-type
methods and some simplified approaches for numerical solution of the ideal
MHD equations. One of the difficulties in the solution of the MHD system is
the satisfaction of the magnetic field divergence-free condition.
Different ways to solve this task are discussed.  If the magnetic field
vector in the uniform interstellar medium flow is not parallel to the
velocity vector, the problem becomes three-dimensional. Both approximate
and exact numerical solutions are considered which were applied in this
case.

Far-field numerical boundary conditions play an essential role in
astrophysical applications owing to very large length scales usual
for these problems. We discuss several approaches that may be useful to
solve problems similar to the stellar wind and interstellar medium
interaction.

\end{abstract}

\section{INTRODUCTION}

The problem of the stellar wind, with the emphasis on the solar wind,
interaction with the interstellar medium has long been the topic of
interest for astrophysicists and specialists in the field of the
solar--terrestrial physics \cite{kn:Alfven}, \cite{kn:Axford},
\cite{kn:Brandt}, \cite{kn:Davis},  and \cite{kn:Parker}. In
\cite{kn:Bar77}, the application of
the continuum equations for this problem is systematically discussed.
The first qualitative model for the interaction of the solar wind~(SW) and
the local interstellar medium~(LISM) was proposed by
E.N.~Parker~\cite{kn:Parker}.
He assumed the interstellar medium to be a subsonic stream with the Mach
number $M_\infty \ll 1$ (see Fig.~1).

Here HP is a heliopause dividing the
SW and the LISM flow. Generally speaking the above assumption is not
correct, since the velocity of the interstellar medium is $V_\infty \sim 20$
km/s and the scattering experiments for the solar radiation show that the
temperature $T_\infty$ of charged particles constituting the LISM is
about $10^{4} K$. Taking into account that the number density of charged
particles is usually considered to be $n_\infty \sim 0.1\,{\rm cm}^{-3}$,
the LISM flow can most likely be supposed supersonic than subsonic. The
supersonic model of the interaction was first proposed in~\cite{kn:Bar70}.
The important feature of this approach lies in the application of the
continuum (Euler gasdynamic) equations only to the charged particles of
both counteracting winds.
Although the presence of turbulent pulsations of plasma is supposed to be
insignificant for the mean flow structure, their influence is realized by a
remarkable change of transport coefficients due to the possibility of
scattering of charged particles on electromagnetic plasma fluctuations.
This results in the substantial decrease of their mean free path compared
with that calculated on the basis of the Coulomb collisions. The solar
wind consists mainly of electrons and protons with the number density
$n_e \sim 10\, {\rm cm}^{-3}$ and velocity $V_e \sim
400\hbox{--}500\rm\,km/s$ and is also considered supersonic. The
index~``$e$'' corresponds to values measured at $1 \AU = 1.5\times
10^{11}\,$km, that is, at the Earth distance from the Sun. Thus, we can
consider this problem, from a gasdynamic viewpoint, as a an interaction of
the supersonic spherically-symmetric (or asymmetric) source flow of SW with
the uniform supersonic LISM flow. This assumption gave rise to a so-called
two-shock model. Generally speaking, this model can be easily obtained
numerically if one choses as initial data for this interaction the
arbitrary jump between the SW and the LISM parameters.

The source flow is essentially a combination of supersonic jet and
blunt-body flows which are perfectly well studied and described in classical
gas dynamics. The extension to the solar wind and space situation
\cite{kn:Dryer}, \cite{kn:Spreiter}
has been validated by numerous spacecraft observations of the plasma
environment of several planets. Different flow regimes and shock-wave
flow structure for the SW--LISM interaction was discussed
in~\cite{kn:Wallis76}.

\begin{figure}[t]
\centerline{\psfig{figure=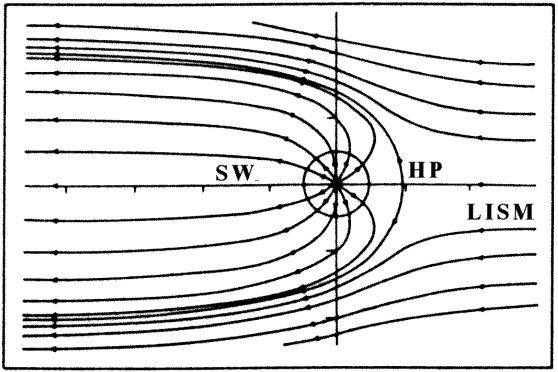,height=5cm}}
\vspace{-6pt}
\caption{ Schematic picture of the SW--LISM interaction~[3]}
\vspace{-12pt}
\end{figure}

In \cite{kn:Bar79} and \cite{kn:Zaitsev}  the axisymmetric problem of the
interaction was considered on the basis of the shock-fitting approach,
but due to the limitations of the applied numerical method the authors
calculated only the upwind part of the
flow. In \cite{kn:Mats89} and \cite{kn:Sawada} the problem of the stellar
wind interaction with the interstellar medium was
analyzed in the closed region surrounding the star. These numerical results
confirmed the scheme~\cite{kn:Wallis76},
but the bullet shape of the internal shock was obtained for a broader range
of parameters. The general schematic picture is shown in Fig.~2.

Here TS
is the inner shock terminating the solar wind within the heliopause (HP),
BS is the bow shock, or the outer shock. TS at a certain point may turn
to form the Mach disk (MD). At this triple point the reflected shock (RS)
and the slip line (SL) originate.

This picture is similar to that suggested for the SW--comet interaction
in~\cite{kn:Wallis76}. The termination shock configuration is caused by its
Mach-type reflection from the symmetry axis.

In \cite{kn:Axford}, \cite{kn:Blum}, \cite{kn:Dessler},
\cite{kn:Fahr73}, \cite{kn:Grz75}, \cite{kn:Wallis71}, and \cite{kn:Wallis75}
the opinion was stated that the resonance recharge processes
between the neutral and the charged particles should strongly influence
the flow picture. The description of
the neutral particle motion cannot be made on the basis of the continuum
approach. For this reason in \cite{kn:Bar82} an approximate method of
taking into account this influence was suggested and, later, in
\cite{kn:Bar93} a self-consistent model was developed that takes into
account the recharge processes. The Monte-Carlo method was used to
calculate the trajectories
of neutral particles. As was admitted in~\cite{kn:Wallis71} and
confirmed by \cite{kn:Bar82} and \cite{kn:Bar93},
the charge-exchange effect effectively diminishes the Mach number of
the LISM flow. This justifies development of alternative subsonic models of
the interaction \cite{kn:Khabib}.

\begin{figure}[t]
\centerline{\psfig{figure=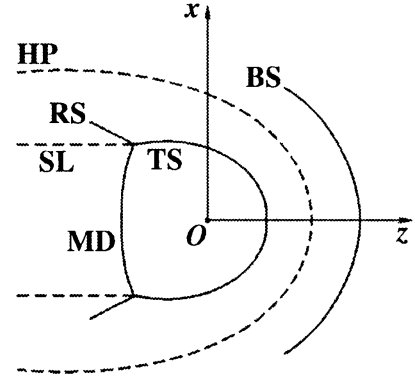,height=5cm}}
\vspace{-6pt}
\caption{General scheme of supersonic interaction}
\vspace{-12pt}
\end{figure}

Of great importance are also nonstationary problems
associated with the variable solar activity. In
\cite{kn:Barnes93}, \cite{kn:Po93}, and \cite{kn:Stein94b} the problem
was investigated  of the time-dependent SW
perturbation influence on the whole flow structure. The nonstationary
picture of the flow and the termination shock response to the 11-year
variation of the solar wind were studied in \cite{kn:Barnes95},
\cite{kn:Karmesin}, and \cite{kn:Po95}.

T. Matsuda et al. \cite{kn:Mats89} discussed the instabilities of the
contact discontinuity dividing the SW and the LISM flow. These
instabilities originated in its lateral region and near the stagnation
point of the flow.
Unstable  solutions were obtained in the parameter range which was not
exactly
suitable for the solar wind and the local interstellar medium flow and,
therefore, were disputed in \cite{kn:Stein94a}. Recently in
\cite{kn:Liewer} a hydrodynamic instability of the heliopause driven by
plasma--neutral charge-exchange processes was discussed.

According to the solar minimum observations by the Ulysses spacecraft,
the solar wind properties depend on helioaltitude. The solar wind in this
case is no longer spherically-symmetric (see also \cite{kn:Suess87}). In
\cite{kn:Pauls96} a 3D problem of the interaction was used and obtained
results were compared with those for the isotropic solar wind.

The influence of the interstellar magnetic field can be important in the
regions, for which the magnetic pressure is comparable  by its value with
the dynamic pressure of the flow. Magnetic field causes  an
increase of the maximum speed of small perturbations in the LISM flow,
thus leading to the decrease of its effective Mach number
\cite{kn:Fahr86}, \cite{kn:Fahr88}.
The ordinary Mach number of the LISM is often assumed to be $M_\infty =
2$.  At the same time the magnitude of
the LISM magnetic field, although not known very well, is estimated
within $7 \times 10^{-7}$ and $3 \times 10^{-6}$ Gauss~\cite{kn:Axford}.
This means, as will be shown later, that the magnetic pressure can exceed
the value of the thermal pressure and is the reason of including
magnetic field into consideration. Its influence can be especially
effective when the charge-exchange processes are included, since both
this effects lead to the decrease of the effective Mach number. This
means that the LISM flow can become subsonic and the bow shock can disappear
(see also \cite{kn:Fahr86} and \cite{kn:Wallis71}). The
stellar wind--interstellar medium interaction with taking into account
magnetic field was first studied  in \cite{kn:Mats93}, although some of the
results seem to be misinterpreted (see~\cite{kn:Bar95}). In the latter
paper the problem was investigated by the shock-fitting method only in the
upwind part of the flow due to the limitations of the numerical scheme.  In
\cite{kn:Po97} the solution was presented of the axisymmetric problem (the
LISM magnetic field strength vector was assumed to be parallel to its
velocity vector) of the SW--LISM interaction in the closed region
surrounding the star. In \cite{kn:Fahr88} the shape of the heliopause was
studied on the basis of the Newtonian approach in the 3D case of the
arbitrary angle between the LISM magnetic field and velocity vectors.
The authors found this shape approximately by equating the values of the
total pressure on the both sides of the heliopause. In~\cite{kn:PoMats}
this problem was first examined numerically.

The complicated pattern of the flow containing a number of interacting
shocks, enhanced by various physical phenomena, makes it a challenging
problem for the application of modern numerical methods invented for pure
gasdynamic and magnetogasdynamic application. Numerical solution of this
problem is often associated with the solution of such accompanying problems
as non-reflecting boundary conditions, numerical implementation of the
condition of the magnetic charge absence, etc. These problems are of
general importance for the young, but quickly developing, field called
computational fluid dynamics. In  Section~2 of this review we present
the mathematical statement of the problem, write out the system of
governing equations and boundary conditions. The choice of initial
conditions for the magnetogasdynamic~(MHD) interaction is discussed.
In Section~3 we discuss stationary solutions of the gasdynamic problem on
the basis of shock-fitting and shock-capturing methods. In Section~4
different approaches are discussed which take into account the
charge-exchange processes. In Section~5 we describe instabilities originating
under certain circumstances in this problem and the reasons causing them.
In Section~6 nonstationary solutions are considered resulting from the
solar wind disturbances and its periodicity. In Section~7 we briefly
discuss the effects of the solar wind spatial asymmetry. And, finally,
Section~8 deals with numerical modeling of the solar wind
interaction with the magnetized interstellar medium.

\section{Mathematical statement of the problem}

To make the paper more concise, we write out in this section the
mathematical statement of the problem based on the MHD equations. The Euler
gasdynamic equations can be easily obtained from the former one by
omitting the  terms containing the magnetic field strength and the equations
describing the behavior of its components.

\subsection{The system of governing equations}

The system of governing equations for a MHD flow of an ideal,
infinitely conducting, perfect plasma in the  Cartesian coordinate
system $x$,~$y$,~$z$, shown in Fig.~2 ($y$-axis is perpendicular to
the picture plane), can be written as follows (one fluid approximation):
\begin{equation}
  {\partial {\bf U}\over \partial t} +
  {\partial {\bf E}\over \partial x} +
  {\partial {\bf F}\over \partial y} +
  {\partial {\bf G}\over \partial z} + {\bf H} = \mathchar"0630\, ,
\end{equation}

\noindent
where
\begin{eqnarray*}
  & &{\bf U} = \left[\begin{array}{c}
      \rho\\ \rho u\\ \rho v\\ \rho w\\ e\\ B_x\\ B_y\\ B_z\\
     \end{array}
    \right] ,\quad
     {\bf E} = \left[\begin{array}{c}
      \rho u\\
      \ds \rho u^{2}+ p_0-{B_x^2 \over 4\pi}\\ \sskip
      \ds \rho uv-{B_xB_y \over 4\pi}\\ \sskip
      \ds \rho uw-{B_xB_z \over 4\pi}\\ \sskip
      \ds (e+p_0)u-{B_x \over \ds 4\pi}({\bf v}\cdot {\bf B})\\
       0\\
       uB_y-vB_x\\
       uB_z-wB_x\\
     \end{array}
     \right] ,\, \\
  & &{\bf F} = \left[\begin{array}{c}
      \rho v\\
      \ds \rho uv-{B_xB_y \over 4\pi}\\
      \ds \rho v^{2}+p_0-{B_y^2 \over 4\pi}\\ \sskip
      \ds \rho vw-{B_yB_z \over 4\pi}\\ \sskip
      \ds (e+p_0)w-{B_y \over 4\pi}({\bf v}\cdot {\bf B)}\\ \sskip
      vB_x-uB_y\\
      0\\
      vB_z-wB_y\\
     \end{array}
      \right] ,\, \\
  & &{\bf G} = \left[\begin{array}{c}
      \rho w\\
      \ds \rho uw-{B_xB_z \over 4\pi}\\ \sskip
      \ds \rho vw-{B_yB_z \over 4\pi}\\
      \ds \rho w^{2}+p_0-{B_z^2 \over 4\pi}\\ \sskip
      \ds (e+p_0)w-{B_z \over 4\pi}({\bf v}\cdot {\bf B)}\\
      wB_x-uB_z\\
      wB_y-vB_z\\
      0\\
    \end{array}
      \right]
\end{eqnarray*}

In system~(1) $\rho$, $u$, $v$, $w$, $B_x$, $B_y$, and $B_z$ are the density
and
the components of the velocity $\bf v$ and of the magnetic field strength
vector $\bf B$. We introduced here also the total pressure $p_0=p+{\bf
B}^2\!/8\pi$ ($p$ is the thermal pressure) and the total energy per unit
volume \[
e={p \over \gamma -1}+{\rho (u^2+v^2+w^2) \over 2}+ {{\bf B}^2 \over
8\pi}\, ,
\]
where $\gamma=5/3$ is the specific heat ratio corresponding to
the fully ionized plasma. The quantities of
density, pressure, velocity, and magnetic field strength are normalized,
respectively, by $\rho_\infty$, $\rho_\infty V^2_\infty$, $V_\infty$, and
$V_\infty \sqrt{\rho_\infty}$, where the index ``$\infty$'' marks the
values in the uniform LISM flow. Time and the linear dimension are
respectively related to $L/V_\infty$ and $L$, where $L$ is equal to
$1\,\AU$. The above formulation implies that molecular and magnetic
viscosities, heat conductivity, and anomalous transport effects are
neglected. The source term~${\bf H}$ can be both of physical and
of geometrical
origin and will be specified separately for each problem. For example, if
the flow is axisymmetric system~(1) can be rewritten in the plane
coordinate system $x,\,z$ as follows:
\begin{equation}
  {\partial {\bf U}\over \partial t} +
  {\partial {\bf E}\over \partial x} +
  {\partial {\bf G}\over \partial z} + {\bf H} = \mathchar"0630\, ,
\end{equation}

\noindent
where
\begin{eqnarray*}
  & &{\bf U} = \left[\begin{array}{c}
      \rho\\ \rho u\\  \rho w\\ e\cr B_x\\  B_z\\
     \end{array}
    \right] ,\quad
    {\bf H} = {1 \over  x}\left[\begin{array}{c}
      \rho u\\
      \ds  \rho u^2 -{B_x^2 \over 4\pi}\\ \sskip
      \ds \rho uw -{B_xB_z \over 4\pi}\\ \sskip
      \ds (e+p_0)u -{B_x \over 4\pi}({\bf v}\cdot {\bf B})\\
       0\\
       uB_z-wB_x\\
     \end{array}
  \right] , \\
  & &{\bf E} = \left[\begin{array}{c}
      \rho u\\
      \ds \rho u^{2}+ p_0-{B_x^2 \over 4\pi}\\ \sskip
      \ds \rho uw-{B_xB_z \over 4\pi}\\ \sskip
      \ds (e+p_0)u-{B_x \over \ds 4\pi}({\bf v}\cdot {\bf B})\\
       0\\
       uB_z-wB_x\\
      \end{array}
  \right] ,\, \\
  & &{\bf G} = \left[\begin{array}{c}
      \rho w\\
      \ds \rho uw-{B_xB_y \over 4\pi}\\
      \ds \rho w^{2}+p_0-{B_z^2 \over 4\pi}\\
      \ds (e+p_0)w-{B_z \over 4\pi}({\bf v}\cdot {\bf B)}\\
      wB_x-uB_z\\
      0\\
     \end{array}
      \right]
\end{eqnarray*}

This system is valid in the half-plane $x0z$ and can be obtained from~(1) in
the assumption of cylindrical symmetry. Its another form is
\begin{equation}
  {\partial {x \bf U}\over \partial t} +
  {\partial {x \bf E}\over \partial x} +
  {\partial {x \bf G}\over \partial z} + {\tilde {\bf H}} = \mathchar"0630\, ,
\end{equation}

\noindent
where
\[
{\tilde{\bf H}} = [0,{-p_0},\, 0,\, 0,\, 0,\, 0]^{\rm T}
\]

Though both presentations of governing equations are mathematically
equivalent, for numerical reasons it is often more convenient to use the
former one, since it is supposed to give more stable results in the
vicinity of the geometrical singularity $x=0$. The Euler gasdynamic
equations can be obtained in different forms from systems~(1) and~(2) by
assuming ${\bf B} \equiv \0$.

\subsection{Initial and boundary conditions}

Calculations are usually performed in the computational region between
the inner and the outer spherical surface (circular in the axisymmetric
case). The flow from the Sun is supposed to be supersonic at the termination
shock distance. For this reason we specify all parameter values at the
inner boundary sphere. The uniform LISM flow is also supersonic and we
can specify the parameter values at the inflow side of the outer boundary.
The treatment of outflow boundary can be more complicated.
In~\cite{kn:Stein94a} all parameters were extrapolated with the zeroth
order along the fluid particle
trajectory. This approach cannot be expected  suitable for deeply
subsonic outflow boundary. Another approach was proposed
in~\cite{kn:Sawada}.
This  method   is   based   on introducing imaginary cells next to the
boundary.  These  cells  are filled with the LISM gas at infinity.
To find the flux through the outer boundary the Riemann problem is solved
between the imaginary and the adjacent cell values. At the boundary segments
with the subsonic--supersonic transition the rarefaction wave relations
are used. This results in the following interpretation (see
\cite{kn:Po93} and \cite{kn:Po96}) of the method initially developed for
a purely gasdynamic case and makes possible its extension to MHD
problems~\cite{kn:Po97}.

Consider the  method based on the two
strictly nonreflecting conditions: the well-known extrapolation condition
for a supersonic exit that provides a characteristically compatible
approximation of the equations on the boundary and the
procedure~\cite{kn:Po93}, \cite{kn:Po96}, developed earlier  for gasdynamic
flows.

The idea of application of the relations in the rarefaction wave for the
realization of the far-field boundary conditions lies in the
artificial locating of the sonic point on the exit boundary. If the flow
is supersonic at infinity such a procedure gives reasonable results and
allows one to perform calculations in the cases for which other
known approaches fail. The interpretation of our method is the
following. Assume that parameters inside the chosen computational region
fully define the flow behavior outside the boundary. In the case of subsonic
exit the only possible elementary Riemann problem configuration for the
above system is a rarefaction wave whose fan covers the boundary.  In
this case, if the self-similar variable value is known, we can locally
continue the internal field to the boundary. That is why, an additional
condition is that the flow velocity attains the sonic value there.

Consider the hyperbolic system for the vector $\bf U$ of unknown
variables in
the  vicinity of the boundary (the right one for definiteness) in the form
\begin{equation}
{\partial {\bf U} \over \partial t} + A {\partial {\bf U} \over \partial x}
= \mathchar"0630\, ,
\end{equation}

\noindent
where ${\bf U}= {\bf U}(x, t)$, and $x$ is the variable in the direction
normal  to the boundary $\Gamma$, $t$ is time, and $A({\bf U})$ is the
coefficient matrix with a complete set of eigenvectors and only
real  eigenvalues. We seek the solution in the form of a simple wave
${\bf U}= {\bf U}(x, t)={\bf U}(\xi)$, where $\xi= {x \over t}$. By
substituting this representation into Eq.~(4), we obtain
\begin{equation}
(A-\lambda I) {\bf U}_\xi = {\bf 0}, \quad \lambda = \xi \, ,
\end{equation}

\noindent
where $I$ is the identity matrix. Owing to Eq.~(5), the vector ${\bf
U}_\xi$ is
the eigenvector of A for the eigenvalue $\lambda =\xi$. This means that we
need to solve the following system of ordinary differential equations
supplemented by the nondifference relation:
\begin{equation}
{\bf U}_\xi = d({\bf U}, \lambda) {\bf r}{(\bf U}, \lambda)\, , \quad
\lambda ({\bf U}) =\xi\, ,
\end{equation}

\noindent
where $\bf r$ is the right eigenvector (the vector-column) of $A$ defined
up to the scalar multiplier $d$. The eigenvalue in the rarefaction wave
varies like $\lambda ({\bf U}) =\xi$. This condition completes the system for
determining ${\bf U}$ and $d$. While realizing this boundary condition we
must integrate Eq.~(6) over $\xi$  from $\xi_0 = \lambda({\bf U}_0)$,
where ${\bf U}_0$ represents the initial subsonic parameters inside the
region, to $\xi=\xi_\Gamma = 0$, that is, to the sonic point. Consider
this approach, first, for pure gas dynamics. Let us choose the vector of
unknowns in Eq.~(4) in the form ${\bf U} = (\rho, u, v, w, a)^T$, where
$\rho$ is the density, $u$ is the velocity vector component normal to
$\Gamma$, $v$ and $w$ are its tangential components, and $a$ is the speed of
sound. The minimum eigenvalue in this case is $\lambda = u - a$ and the
related eigenvector is
\begin{equation}
{\bf r}= \left[1,\ -{a \over \rho},\ 0,\ 0,\ {(\gamma -1) a \over 2 \rho}
\right]^{\rm T}\, ,
\end{equation}

\noindent
where $\gamma$  is the adiabatic index. System~(6) in this case acquires
the form
\begin{eqnarray}
& &\rho_\xi= d, \ u_\xi = - {ad \over \rho}, \  v_\xi = 0,\ w_\xi=0,
\nonumber  \\
& & \\
& &a_\xi={(\gamma -1)ad \over 2\rho},\ u-a=\xi. \nonumber
\end{eqnarray}

This system can be exactly integrated, as its invariants are
\begin{equation}
\left({p \over \rho^\gamma}\right)_\xi =0,\ \left(u+{2a \over \gamma
-1}\right)_\xi =0, \ v_\xi =0, \ w_\xi =0.
\end{equation}

Thus, we obtain
\begin{eqnarray}
  & &a_\Gamma = {\gamma -1 \over \gamma +1}\left(u_0+{2 \over \gamma -1}
     a_0\right),\ u_\Gamma=a_\Gamma, \ v_\Gamma =v_0 \, , \nonumber \\
  & & \\
  & &w_\Gamma =w_0, \ \rho_\Gamma = \rho_0 \left(a_\Gamma \over
     a_0\right)^{2 \over \gamma -1} \nonumber
\end{eqnarray}

The index ``0'' indicates the values belonging to the inner region. On
the discreet mesh this means that they are taken from the center (or
from the left side) of the cell adjacent to the boundary.
Equations~(10) must be supplemented by the condition at the supersonic
exit if $(u/a)_0 \geq 1$: ${\bf U}_\Gamma = {\bf U}_0$.

These conditions are mutually consistent and coincide for $u_0 =
a_0$. Note that in this case we did not need the explicit
expression for $d$ which can be easily found from the second and the fifth
equations in~(8). Besides the exact derivation based on the relations in
the rarefaction wave, we give here the approximate relations keeping in
mind  such systems for which no exact expressions of this kind can be
written out. For this purpose we first exclude $d$ by substituting the
first equation from~(8) into the other ones. Then we obtain
\begin{eqnarray}
  & &u_\xi=-{a \rho_\xi \over \rho},\ \ v_\xi=0,\ \ w_\xi=0, \nonumber \\
\noalign{\kern -6pt}
  & & \\
\noalign{\kern -6pt}
  & &a_\xi={(\gamma -1)a\rho_\xi \over 2\rho},\ \ u-a=\xi \nonumber
\end{eqnarray}

Now, approximating Eqs.~(11) by finite differences
we arrive at the following relations:
\begin{eqnarray}
  & &a_\Gamma={\gamma -1 \over \gamma +1}\left(u_0+{2\over \gamma -1}
     a_0\right),\ u_\Gamma=a_\Gamma,\ v_\Gamma=v_0\, ,\nonumber \\
  & & \\
  & &w_\Gamma=w_0,\ \rho_\Gamma=\left[1+{2 \over \gamma -1}\left({a_\Gamma
     \over a_0}-1\right)\right]. \nonumber
\end{eqnarray}

In this approximation solution (12) differs from (10)
only in the entropy invariant and represents its linearization.

Now we proceed to the MHD equations. Let us choose
the unknown vector in Eq.~(4) in the form
\begin{equation}
{\bf U} = [\rho,\ u,\ v,\ w,\ a,\ B_x,\ B_y,\  B_z]^{\rm T}.
\end{equation}

\noindent
where, in addition to the purely gasdynamic case, the components appear
of the magnetic field strength vector normal ($B_x$) and tangential ($B_y$
and $B_z$) to $\Gamma$. The minimum eigenvalue in this case is
$\lambda = u - a_f$, where $a_f$ is the largest of the two magnetosonic
speeds $a_f$ and $a_s$ ($a_f > a > a_s$):
\begin{eqnarray}
   & &a_{f,s}=\frac {1}{2} \Biggl[ \left(a^2+{|{\bf B}|^2 \over 4\pi\rho} +
      {a |B_x| \over \sqrt{\pi\rho}}\right)^{1/2}  \nonumber \\
   & & \\
   & &\pm \left(a^2+{|{\bf B}|^2 \over 4\pi\rho} -{a |B_x| \over
       \sqrt{\pi\rho}}\right)^{1/2}\Biggr],\ |{\bf B}|^2 =
       B_x^2+B_y^2+B_z^2 \nonumber
\end{eqnarray}

The eigenvector corresponding to this eigenvalue is
\begin{eqnarray}
  & &{\bf r}=\left[1,\,-{a_f \over \rho},\,\alpha B_y,\,\alpha B_z,\,
     {(\gamma -1)a \over 2\rho},\,0,\,\beta B_y,\,\beta B_z\right]^{\rm T},\,
     \nonumber \\
  & & \\
  & &\alpha ={aa_s \over 2\rho\sqrt{\pi \rho}(a^2-a_s^2)},\,
      \beta={a^2 \over \rho (a^2-a^2_s)}\, \nonumber
\end{eqnarray}

In this case system~(6) acquires the form
\begin{eqnarray}
  & &\rho_\xi=d,\,u_\xi=-{a_f d \over \rho},\,v_\xi=\alpha B_y
      d,\,w_\xi=\alpha B_x d\, , \nonumber \\
  & &a_\xi={(\gamma -1)ad \over 2\rho},\,(B_x)_\xi=0,\,(B_y)_\xi=\beta B_y
     d\, , \\
  & &(B_z)_\xi =\beta B_y d,\, u-a_f=\xi\,  \nonumber
\end{eqnarray}

Now in system~(16) we substitute the equation for $a$ by the equation for
$a_f$, which can be easily obtained from Eq.~(14) by direct differencing
with respect to $\xi$  and by using Eqs.~(16):
\begin{equation}
(a_f)_\xi={\vartheta d \over \rho},\ \vartheta=\rho {\bf r}
{\partial a_f \over \partial {\bf U}}\, .
\end{equation}

It can be shown in this connection that for any admissible
 values of functions we have $\vartheta +a_f \geq 0$.

Then, by passing from $d$ to $\rho_\xi$, we obtain the reduced
system of equations that can be approximated similarly to Eq.~(11)
\begin{eqnarray}
  & &(a_f)_\Gamma=\left({u \vartheta + a^2_f \over \vartheta
     +a_f}\right)_0,\, \rho_\Gamma=\rho_0\left(1+{u-a_f \over \vartheta
     +a_f}\right)_0\, , \nonumber \\
  & &u_\Gamma=(a_f)_\Gamma,\,v_\Gamma=v_0+(\alpha B_y)_0(\rho_\Gamma
     -\rho_0)\, , \nonumber \\
  & &w_\Gamma=w_0+(\alpha B_z)_0(\rho_\Gamma -\rho_0),\,(B_x)_\Gamma=
     (B_x)_0\, , \\
  & &(B_y)_\Gamma=(B_y)_0[1+\beta_0 (\rho_\Gamma -\rho_0)]\, ,\nonumber  \\
  & &(B_z)_\Gamma=(B_z)_0[1+\beta_0 (\rho_\Gamma -\rho_0)]\, ,\nonumber  \\
  & &(a)_\Gamma=(a)_0+(\gamma -1)(a/2\rho)_0 (\rho_\Gamma -\rho_0)\,\nonumber
\end{eqnarray}

Note that in contrast to the purely gasdynamic case the velocity
components tangential to the boundary, generally speaking, are different
from  their internal values in the presence of the magnetic field.

The case of the triple degeneration of eigenvalues
($B_y^2 + B_z^2 = 0$ and $a^2 \to  B_x^2/4\pi \rho$), when $\vartheta +a_f
\to \infty$, can be easily avoided by assuming  $B_y^2 + B_z^2 =
\varepsilon$, where $\varepsilon$ is a small positive number.

The approach described above turned out to give stable results in contrast
to the attempts of a straight application of the well-known non-reflecting
boundary conditions, say, \cite{kn:Thompson}. It is worth mentioning in
this connection that a comprehensive review (more than 200 references)
of various versions and modifications of non-reflecting boundary
conditions can be found in \cite{kn:Ilg85} and \cite{kn:Ilg90} (see also
\cite{kn:Givoli} and \cite{kn:Thompson}).

As initial values  the jump can be chosen between the SW  and the LISM
parameters at a fixed distance $R_{\rm f}$ from the Sun smaller
than the TS stand-off distance. For $R<R_{\rm f}$  the SW
parameter distribution is specified. The  magnetic field
pressure in this region is supposed to be negligibly small
comparing with the SW hydrodynamic pressure, thus ${\bf  B}_e = {\bf 0}$.
For $R>R_{\rm f}$ the uniform distribution of the LISM pressure
and density is assumed. If we solve an MHD problem, that is, the LISM flow
is magnetized, it is wise to specify a magnetic field strength distribution
satisfying the divergence-free condition. For this reason
the magnetic and the velocity field  in the LISM and in the SW flow  are
initially joined in the computational region so that ${\bf B}$ conserves a
constant angle with
${\bf v}$ and $\div {\bf B}=0$. This is done by assuming for $R>R_{\rm f}$
\begin{eqnarray}
  & &U=- V_{\infty} \left[1- \left({R_{\rm f} \over
     R}\right)^3\right] \cos \theta \nonumber \\
  & &V=0 \\
  & &W= V_{\infty}\left[1+ {1\over 2} \left({R_{\rm f} \over R}\right)^3
     \right]\sin \theta\,  \nonumber
\end{eqnarray}

Here $U$, $V$, and $W$ represent the spherical components of the velocity
vector in the directions $R$, $\phi$, and $\theta$, respectively. This
distribution corresponds to an incompressible fluid flow velocity
distribution over a sphere, directed along the $z$-axis ($\theta$-axis).
The magnetic field is initialized in the same way, except that the field
configuration is rotated about the $y$-axis  so
that the magnetic field vector  is tilted with respect to the velocity
vector by the desired angle.

If neutral particles are to be taken into account, we must specify their
number density and velocity at the inflow.

\section{Stationary gasdynamic calculations}

A gasdynamic model for the SW interaction with the supersonic
interstellar wind was first suggested in~\cite{kn:Bar79}. A cylindrical
formulation was adopted which corresponds to a uniform LISM flow and a
spherically-symmetric SW. The calculations were performed on the basis of
a simplified thin-layer approximation. The solar wind was assumed to be
decelerated mainly in the process of its interaction with the charged
component, or plasma component, of the LISM. With the latter assumption
calculations can also be made without simplifications using the Euler
gasdynamic equations. Both shock-fitting and shock-capturing approaches
can be found in publications and we describe them briefly in the following
subsections.
\subsection{Shock-fitting methods}

Taking into account the adopted two-shock model with the contact
discontinuity between the shocks, one can easily use shock-fitting methods in
the upwind part of the interaction region. The idea of this approach lies
in the subdivision of the computational region into subregions of
smooth flow. In these smooth subregions any numerical scheme of
sufficient order of accuracy can be used. Derivatives in this approach
must never be approximated by finite differences across discontinuities.
The latter are traced as boundary lines. Proper jump relations are used to
determine the change of parameters across these boundaries and their new
position in the course of time. This approach is very economical, since
1) you need not perform calculations in the regions of the uniform LISM
and the spherically-symmetric SW, which are known beforehand, thus
reducing the size of the computational domain and 2) you can avoid
spurious oscillations around discontinuities inherent in application of
shock-capturing methods. For this  reason one can use linear high-order of
accuracy numerical schemes not worrying about high non-oscillatory
resolution of discontinuities. In the shock-fitting method their position
and intensity are determined exactly. Certain difficulties originate if
two or more discontinuities are interacting with each other. In this
case, in principle, one can still use a shock-fitting approach using
exact solutions for the problems of a discontinuity interaction. The
algorithm, however, can become rather complicated (see
\cite{kn:Moretti79}, \cite{kn:Moretti84}, \cite{kn:Moretti87},
and \cite{kn:Napolitano}).
In the shock-fitting approach a quasi-linear form of the system of
governing equations is usually chosen rather than a conservation-law form.

On the basis of the shock-fitting approach the problem under
consideration was first solved in~\cite{kn:Bar79} by the
Babenko--Rusanov implicit
scheme~\cite{kn:Babenko}. Due to the limitations of the numerical approach
only upwind part of the interaction was calculated. A polar coordinate
system $R,\, \theta$ was used and calculations were performed in the
region restricted by the ray $\theta = \theta_{\rm max}$ at which the
velocity normal to this boundary remained supersonic. The system of
linear algebraic equations on the computational grid was solved
together with the Rankine--Hugoniot conservation relations and the
relations on the contact discontinuity. The computational region was
located between the termination  and the bow shock. The
steady-state solution was obtained as $t \to \infty$ with the boundary
conditions independent of time.

\begin{figure}[t]
\centerline{\psfig{figure=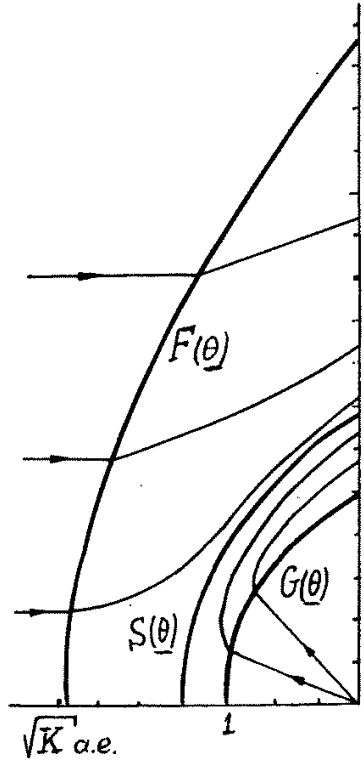,height=10.16cm}}
\vspace{-6pt}
\caption{Self-similar picture of shock wave position [7]}
\vspace{-12pt}
\end{figure}

\begin{figure}[t]
\centerline{\psfig{figure=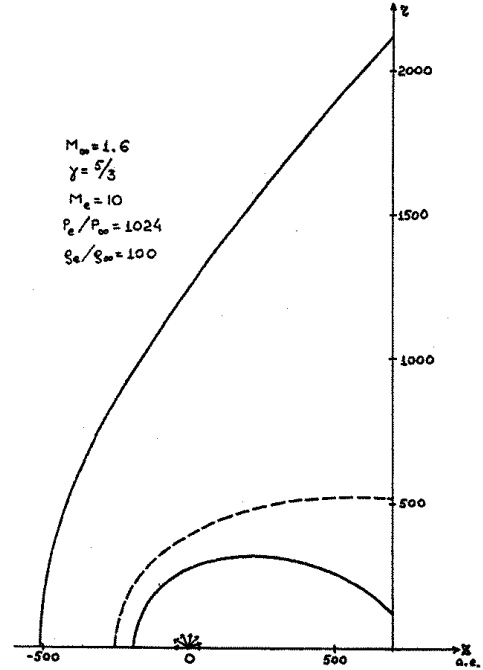,height=8.89cm}}
\vspace{-6pt}
\caption{ Discontinuity pattern: shock-fitting approach [113]}
\vspace{-12pt}
\end{figure}

A physicist usually seeks dimensionless similarity parameters of the
problem. For the considered problem they are represented by the Mach
numbers $M_e$ and $M_\infty$ of the SW and the LISM flow, respectively, the
ratios of their dynamic pressures $K=n_e V_e^2/n_\infty V_\infty^2$ and
stagnation temperatures $\chi= T_{e0}/ T_{\infty 0}$. The specific heat
ratio $\gamma$ is usually adopted to be equal to 5/3 which corresponds to
the fully ionized plasma. If neutral particles are not taken into account,
the flow from the inner side of TS becomes hypersonic and, therefore,
the results are only weakly dependent on the choice of $M_e$ taken at 1
\AU. Though results formally depend on $\chi$ (say, for $\chi =1$ the
density jump over CD is absent) an analysis of
conservation relations on the discontinuities in the upwind part of
the interaction region make us conclude that they can be recalculated
using results for one particular value of $\chi$. As $K$ is the
similarity parameter (see \cite{kn:Bar70} and \cite{kn:Bar79}),
for the hypersonic solar wind
and $K > 1$ the results are independent  of $K$ if the distances are
measured in units of $1 \AU \times \sqrt K$. In Fig.~3, the shapes of
the outer (F) and of the inner (G) shock wave, the contact surface (S) and
some streamlines are presented for the case $M_\infty =2$.

In~\cite{kn:Zaitsev} a new shock-fitting numerical algorithm was described
and applied to the problem under consideration that provided results
with such a high accuracy that in the upwind part of the interaction it
may represent a testing benchmark for all newly  developed methods.
This method is based on the composite explicit-implicit finite-difference
scheme~\cite{kn:Radvogin}. The well-known explicit Lax--Wendroff and the
implicit Babenko scheme~\cite{kn:Babenko} represent its constituent parts.
The algorithm is organized in such a way that depending on the CFL
number value either explicit or implicit approximation is used.
This approach makes the proposed method more economical than
a purely implicit scheme. Among the schemes based on this
approach we can also mention \cite{kn:Mac} and \cite{kn:Po85}. The latter
method is the extension of MacCormack's explicit-implicit scheme for
the steady Euler equations hyperbolic with respect to one of the
space coordinates. A steady-state solution of the SW--LISM interaction
was obtained in~\cite{kn:Zaitsev} by a quasi-marching method in which
nonstationary problem was solved for $t\to\infty$ for each
radial ray. Such an approach also allowed to save a
computational time comparing with a direct solution of
the nonstationary system. One of the computational results is
shown in Fig~4.

One must admit here that, in contrast to the results from \cite{kn:Bar79},
calculations were performed for a rather long distance in the downwind
direction, actually
until the flow along the $z$-axis shown in Fig.~4 remained supersonic.
This resulted in a considerable unjustifiable elongation of the
termination shock, as will be seen from the subsequent subsection.
The discrepancy between the results is caused by the
limitation of the quasi-marching approach that  cannot take into account
the Mach-type reflection of the inner shock from the symmetry axis.
Nevertheless, the results are without any doubt quite reliable up to a
certain distance in the wake region. No need to mention that the time
necessary for obtaining the steady-state solution in this case is
considerably less than that in the case of applying  shock-capturing
methods.

\subsection{Shock-capturing methods}

In shock-capturing methods we calculate finite differences across
discontinuities. This may cause spurious oscillations of the solution if
non-monotone numerical schemes are used. On the other hand, all linear
schemes of the order of accuracy higher than one are
non-monotone~\cite{kn:Godunov}. For
this reason one or another artificial viscosity must be
used~\cite{kn:Pulliam} or
nonlinear numerical schemes ought to be applied. It is not our task to give a
review of high-resolution TVD (total variation diminishing) schemes in
this paper (see \cite{kn:Hirsch} and \cite{kn:Yee} for a regular
mathematical background).
Although both finite-difference and finite-volume methods can be
equivalently applied in the latter schemes, we shall dwell mainly on the
finite-volume formulation  and monotonic upstream schemes for conservation
laws (MUSCL) approach, since they are more descriptive.

To solve axisymmetric system~(2), let us introduce a polar mesh
\begin{eqnarray}
  & &f^k_{l,n}=f(R_l, \theta_n, t^k)\, ,\quad t^k=k \Delta t\, ;\nonumber \\
  & & R_{l} = R_{\rm min} + (l-1)\Delta R,\quad l=1,2,\ldots,L; \nonumber \\
  & & \theta _{n} = (n-2.5) \Delta \theta,\quad n=1,2,\ldots,N; \\
  & & \Delta R = (R_{\rm max}- R_{\rm min})/(L-1), \nonumber\\
  & & \Delta \theta = \pi /(N-4)\nonumber
\end{eqnarray}

\noindent
with the center in the star. Then for each cell system (2) in the
finite-volume formulation can be rewritten as follows:
\begin{eqnarray}
   & &R_{l}\,\Delta R\,\Delta\theta\,{{\bf U}_{l,n}^{k+1} -
     {\bf U}_{l,n}^{k} \over \Delta t}+{}
 \nonumber  \\
   & &(R_{l+1/2}\bar {\bf E}_{l+1/2,n} +
    R_{l-1/2}\bar {\bf E}_{l-1/2,n})\, \Delta \theta+{}
  \\
   & &(\bar {\bf E}_{l,n+1/2} +\bar {\bf E}_{l,n-1/2})\, \Delta R +
    R_{l}\, \Delta R\, \Delta \theta\,  {\bf H}_{l,n} = 0 .
  \nonumber
\end{eqnarray}

Here $\bar{\bf E}$ is the flux normal to the boundary, defined as:
\begin{equation}
  \bar{\bf E} = n_1 {\bf E} + n_2{\bf G},
\end{equation}

\noindent
where ${\bf n}= (n_{1},n_{2})$ is a unit outward vector normal
to the cell surface.

Equation (21) has a time-discretized conservation-law form for an individual
computational cell. Various numerical schemes are specified by the
method chosen to  calculate the numerical flux $\bar{\bf E}$ through the
cell boundary surfaces. In~\cite{kn:Stein94a} the two-step Lax--Wendroff
scheme is used with the second order of accuracy. As  usual for such
schemes, an additional smoothing must be introduced to remove
high-frequency oscillations and  overshoots and undershoots
originating near smeared shocks owing
to the non-monotonicity of the scheme. Another possible flux calculation
formulas which also include artificial viscosity  were  applied
in~\cite{kn:Pauls95}. It is based on the ZEUS fractional step code
\cite{kn:Stone}.
All methods using artificial viscosity for oscillation damping contain an
empirical viscosity coefficient which must be adjusted in a way suitable
for any particular problem. The compromise is between the effective
monotonization of the solution and its deterioration. Nonlinear
high-resolution numerical schemes are free from this drawback.

To attain the second order of accuracy in space, a piecewise-linear
distribution of parameters inside computational cells can be
adopted~\cite{kn:Yee}.
One can use the simplest ``minmod'' reconstruction procedure
\begin{eqnarray}
  & &{\bf U}^R_{l+1/2} = {\bf U}^{k}_{l+1} - { \frac{1}{2}} \minmod
   (\Delta {\bf U}^k_{l+1/2}, \Delta {\bf U}^k_{l+3/2})\, , \\
  & &{\bf U}^L_{l+1/2} = {\bf U}^{k}_{l} + { \frac{1}{2}} \minmod (\Delta
  {\bf U}^k_{l-1/2}, \Delta {\bf U}^k_{l+1/2})\, , \\
  & &\minmod(x,y) = \sgn (x) \, \max \{0, \min [|x|, y \sgn (x)]\}\, ,
\nonumber
\end{eqnarray}

\noindent
where $\Delta {\bf U}^k_{l+1/2} = {\bf U}^{k}_{l+1} - {\bf U}^{k}_l$,  and
${\bf U}^R_{l+1/2}$ and ${\bf U}^L_{l+1/2}$ defined by Eqs.~(23)--(24)
represent parameter values on the right and on the left side  of the cell
surface with the index ``$l+1/2$''. The index~``$n$'' is omitted in these
formulas. The reconstruction procedure in the angular direction is similar.
To attain better resolution of the contact discontinuity, one can use more
compressive slope limiting procedure for the density, e.~g.,
\begin{eqnarray}
& &\rho^R_{l+1/2}= \rho^k_{l+1}- \minmod (\Delta \rho^k_{l+1/2},
 \Delta \rho^k_{l+3/2}, {\tilde \Delta}) \, , \\
& &\rho^L_{l+1/2}= \rho^k_{l}+ \minmod (\Delta \rho^k_{l-1/2},
 \Delta \rho^k_{l+1/2}, {\tilde {\tilde \Delta}}) \, , \\
& &{\tilde \Delta}= 0.25 (\Delta \rho^k_{l+1/2}+ \Delta \rho^k_{l+3/2})\, ,
\nonumber \\
& &{\tilde {\tilde \Delta}}= 0.25 (\Delta \rho^k_{l-1/2}+ \Delta
\rho^k_{l+1/2})\, , \nonumber \\
& & \minmod(x,y,z) = \sgn (x) \, \max \{0, \min [|x|, y \sgn (x),
   |z|] \} \nonumber
\end{eqnarray}

The fluxes $\bar {\bf E}({\bf U}^R,{\bf U}^L)$  through the cell surfaces can
be found by different methods. Wide recognition acquired TVD
shock-capturing methods based on the exact or some of the approximate
solutions to the Riemann problem. The first steady-state solution of the
SW--LISM interaction problem in the closed region surrounding the star was
obtained in~\cite{kn:Sawada} on the basis of the Osher
approximate Riemann problem solver~\cite{kn:Chak}. In this approach an
approximate solution to the Riemann problem is formed using elementary
simple waves which correspond to definite eigenvectors of the Euler
gasdynamic system and separate the regions of the constant flow. In the
exact solution, a rarefaction wave, a contact discontinuity and a shock
wave generally appear in various combinations. In the approximate
solution used in~\cite{kn:Sawada}
shock waves are approximated by compression waves. This approach connects
parameter values on the right and on the left side of the computational cell
by a number of algebraic relations. Such approach, however, seems to be
more time-consuming than that based on Roe's solution of the linearized
Riemann problem~\cite{kn:Roe81} (see also a comprehensive description of
characteristic-based schemes for the Euler equations~\cite{kn:Roe86}).
A numerical flux in this case is calculated as follows:
\begin{equation}
\bar {\bf E}({\bf U}^R,{\bf U}^L)={1\over 2}
\left[{\bf E}({\bf U}^L) + {\bf E}({\bf U}^R) -S|\Lambda|S^{-1}({\bf U}^R -
{\bf U}^L)\right]
\end{equation}

Here $S(\bar{\bf U})$ and $S^{-1}(\bar{\bf U})$ are the matrices formed by
the right and by the left eigenvectors, respectively, of the frozen
Jacobian matrix
\[
  \bar J = {\partial \bar{\bf E}(\bar{\bf U})\over \partial {\bf U}}
\]

The value of $\bar{\bf U}({\bf U}^L, {\bf U}^R)$ is chosen so that the
conservation relations on shocks are exactly satisfied. The matrix $|\Lambda|$
is a diagonal matrix consisting of the frozen Jacobian matrix eigenvalue
moduli.

The important peculiarity of the latter method is that, although it gives
the solution of the linearized problem, the exact satisfaction of the
Rankine--Hugoniot relations on shocks provides their
more adequate and sharp resolution. This method was applied to the
problem under consideration in \cite{kn:Mats89} and \cite{kn:Po93}.

\begin{figure}[t]
\centerline{\psfig{figure=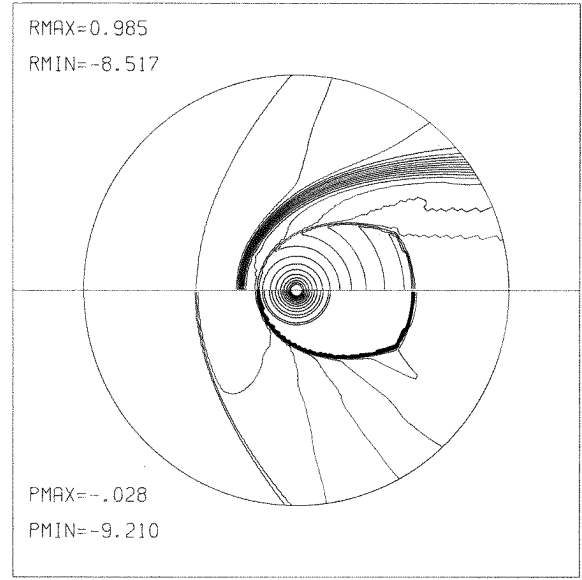,height=7.7cm}}
\vspace{-6pt}
\caption{Discontinuity pattern: shock-capturing approach [75]
}
\vspace{-12pt}
\end{figure}

Although all the mentioned methods give essentially
similar pattern of discontinuities in the computational region, we
present here, for the convenience of the further discussion, the
picture from~\cite{kn:Po93}. The interstellar plasma number density of
protons is assumed to be
$n_{H+} \approx \rm1\, cm^{-3}$.  The velocity of
LISM relative to  the solar system is about $\rm20\, km/s$, while the
speed of sound of  the  LISM gas is about $\rm10\,km/s$. Thus, LISM flow
is supersonic. The SW charged particle number density is chosen
$n_{H+}\approx \rm10\, cm^{-3}$, its velocity is
$V_{e} \approx \rm500\, km/s$, the speed of sound $c_{e}\approx \rm100\,
km/s$ at the distance of  the Earth's orbit ($1\,\AU$).
The nondimensional parameters of the  problem  are  Mach  numbers  of
LISM and  SW $M_{\infty }$, $M_{e}$, the relations of dynamic pressures
$K= \rho _{e}V^{2}_{e}/\rho _{\infty }V^{2}_{\infty }$
and stagnation temperatures $\chi= T_{0e}/ T_{0\infty }$  of SW  and  LISM.
This corresponds to the following values of dimensionless parameters of
the problem: $M_{\infty }=2$, $M_{e}=5$, $\chi  =400$, $K = 6250$.
The specific heat ratios for SW and  LISM  are
supposed to be~5/3. The calculation is performed in the  ring  region
with the inner and outer circle radii being $R_{\rm min }= 10$
and $R_{\rm max }= 500\,\AU$ with 99 and 116 cells in the radial and in
the angular direction, respectively.
Constant pressure (below the symmetry axis) and constant  density natural
logarithm  contours  of  the steady-state solution are presented  in
Fig.~5. Results are given in the polar region with the  inner  and
outer  circle radii 10 and~$400\,\AU$. All peculiarities are seen  of
the  shock wave pattern shown schematically in Fig.~2.

\begin{figure}[t]
\centerline{\psfig{figure=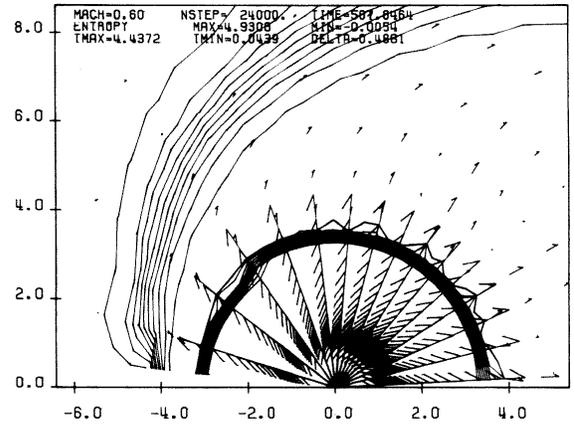,height=5.7cm}}
\vspace{-6pt}
\caption{Entropy contours with velocity vectors for the model with
the Mach number 0.6 [93]
}
\vspace{-12pt}
\end{figure}

\begin{figure}[t]
\centerline{\psfig{figure=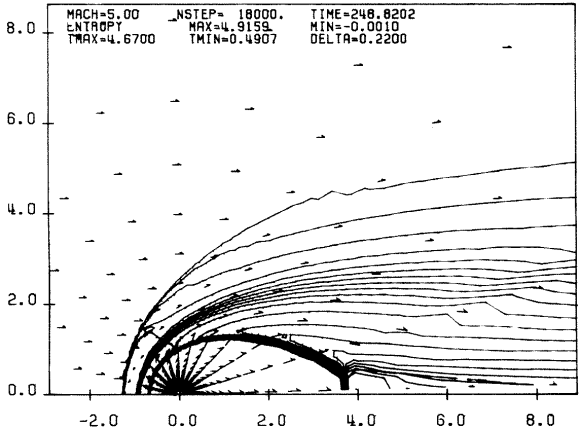,height=5.7cm}}
\vspace{-6pt}
\caption{Entropy contours and velocity vectors for the Mach number 5 [93]
}
\vspace{-12pt}
\end{figure}

\begin{figure}[t]
\centerline{\psfig{figure=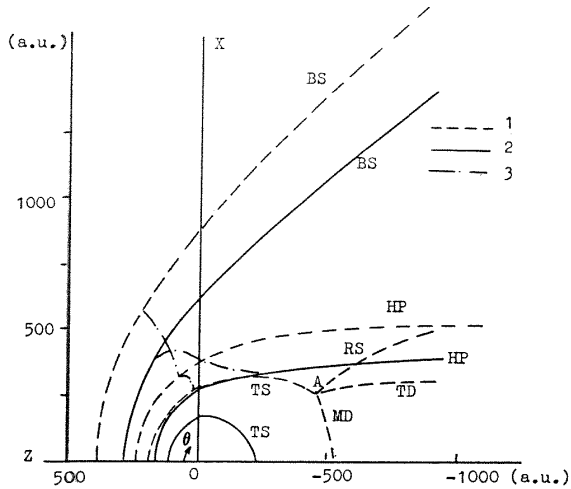,height=6.7cm}}
\vspace{-6pt}
\caption{
Geometrical pattern of the interface. Results of the numerical calculations
for $n_{H\infty}=0(1)$ and $n_{H\infty}=\rm0.14 cm^{-3}(2)$; curves (3)
are the
sonic lines. Positions of the bow shock (BS), termination shock (TS),
heliopause (HP), reflected shock (RS), tangential discontinuity (TD), and
Mach disk (MD) are shown.
}
\vspace{-12pt}
\end{figure}

Calculations for a large variety of the stellar wind and interstellar medium
parameters, including those very different from the parameters of the
solar wind, were performed in~\cite{kn:Mats89} and \cite{kn:Sawada}. These
parametric studies may
reflect various situations corresponding to ejecting stars and their
environment. Although from the purely gasdynamic viewpoint the LISM flow is
supersonic, the presence of the interstellar magnetic field and
charge-exchange processes can decrease the effective Mach number of the
uniform interstellar medium flow, thus making it subsonic. For this reason
several authors~\cite{kn:Shima}, \cite{kn:Stein94b} performed calculations
of a subsonic interaction on the basis of the Euler equations. It is quite
clear that, unlike  the two-shock Baranov's model, no bow shock can
appear in the stationary solution. As will be shown later, the
termination shock in
this case does not have a bullet shape and the difference between
its stand-off distances in the upwind and in the downwind direction is
not so pronounced. This can also be seen in Fig.~6, corresponding to
$M_\infty=0.6$. If the stellar wind Mach number is larger than that of the
solar wind, the termination shock becomes elongated in the backward direction
and the size of the Mach disk diminishes (see Fig~7, corresponding to
$M_\infty=5$). The latter two figures are taken from \cite{kn:Shima}.

Summarizing the subject of this section, we would like to admit that,
although application of shock-fitting methods can provide a solution
with a very high precision at low computational costs, they can
be used only in the flow regions with a simple shock-wave pattern which
is known beforehand (say, the upwind part of the SW--LISM interaction).
Attempts to promote calculations for larger distances into the downwind
region results in the substantial distortion of the results, like it was
in~\cite{kn:Zaitsev}, in which the possibility of the Mach-type
reflection of the termination
shock was not taken into account. A shock-fitting method in its rigorous
sense was not realized, since the exact relations connecting parameters
in the triple point were not used. The best way out in this case is to
combine shock-fitting and shock-capturing methods, as it was done
in~\cite{kn:Bar93}.

\section{Stationary solutions including neutral particles}

Since we restrict ourselves in this review to the description of numerical
methods used to solve the SW--LISM interaction problem, more or less
complete description of physical processes governing the charge exchange
between the plasma and the neutral component of the flow lies beyond the
scope of this work. It
can be found in the corresponding references mentioned in Introduction.
However, it is quite clear that the LISM is a partially ionized gas and,
therefore, a consistent model must be developed accounting for the mutual
influence of charged and neutral particles. Although both helium and
hydrogen atoms are present, the authors usually disregard helium atoms,
since their cosmic abundance is much less than that of hydrogen atoms.

The first attempts to estimate the influence of this process,
say,~\cite{kn:Bar79} and \cite{kn:Ripken}, did not include the influence of
the hydrogen atoms on the plasma component. The first self-consistent
model of the heliospheric interface was suggested in~\cite{kn:Bar82}. In
that paper the authors introduced the source terms accounting for the
momentum and the energy transfer between the two
components into the system of the Euler gasdynamic equations in the
quasi-linear form. These source terms were presented in the form:
\begin{equation}
\rho \nu_c ({\bf v}_H - {\bf v})
\end{equation}

\noindent
for the momentum equations~(1) and
\begin{equation}
\rho (\gamma -1) \nu_c \left[{({\bf v}_H - {\bf v})^2\over 2} -
{3kT \over 2m_H}+{3kT_H \over 2m_H}\right]
\end{equation}

\noindent
for the energy equation. Here $k$ is the Boltzmann constant and $m_H$ is
the mass of the hydrogen atom. For the collision frequency between
protons and atoms the following formula was used:
$$
\nu_c={\rho_H \sigma Q \over m_H},\,
Q=\left[({\bf v}_H - {\bf v})^2 + {128k(T+T_H) \over
9\pi m_H}\right]^{1\over 2},
$$
where $\sigma$ is the effective charge-exchange cross-section.

The hydrogen atoms were assumed to conserve their velocity and temperature
in the process of their interaction with protons
and the origin of the secondary hydrogen atoms was neglected.
That allowed to describe the behavior of neutrals in the region between the
two shocks by the system
\begin{eqnarray*}
& &{\bf v}_H={\bf v}_{H_\infty}={\rm const},\\
& &T_H=T_{H_\infty}={\rm const}, \\
& &\div(\rho_H {\bf v}_H)=- \rho \nu_c
\end{eqnarray*}

Incorporating this source term into implicit method~\cite{kn:Babenko} does
not represent any difficulty and the solution was obtained for  various
values of dimensionless parameters. The major effect of charge exchange
on the heliospheric interfaces is to decrease the distances to the TS,
HP, and BS.
The original developers of the approach~\cite{kn:Bar82}, which  considers
the flow of neutral particles as a hydrodynamic flow, themselves admitted
in~\cite{kn:Bar90} its main
drawbacks: (1) such description is hardly justifiable, since the mean
free-path of the hydrogen atoms is not smaller than the characteristic
length of the problem; (2) the Maxwellian distribution of the atoms was
adopted to calculate the plasma momentum and energy losses. This was the
reason of applying the Monte-Carlo method for simulation of the hydrogen
atom trajectories~\cite{kn:Malama}. It gives a possibility to evaluate
the source terms in the momentum and the energy equations for the plasma
component on the basis of the kinetic description of the hydrogen atoms.
An iteration method~\cite{kn:Bar91} was proposed to solve both systems of
equations. The numerical results are discussed in~\cite{kn:Bar93} and
\cite{kn:Bar95}. This approach is nowadays the only one which can be
considered physically consistent from the viewpoint of computational fluid
dynamics. In Fig.~8, the geometrical interface between the two flows is
presented~\cite{kn:Bar93} reflecting the influence of the charge exchange
processes. The picture is presented in the $X0Z$ plane, where $0Z$
coincides with the axis of symmetry and is antiparallel to the vector on
the LISM velocity (the Sun is in the coordinate system origin).
The solid and the dotted lines in Fig.~8 correspond to
$n_{H_\infty}$= $\rm0.14\,cm^{-3}$ and $n_{H_\infty}=0$, respectively.
The parameters of the plasma component were the following:
$n_e=\rm7\,cm^{-3}$,
$V_e=\rm450\,km/s$, $M_e=10$, $n_\infty=\rm 0.07\,cm^{-3}$, $V_\infty=
\rm25\, km/s$, and
$M_\infty=2$. One can see a large influence of the described processes on
the flow pattern and it is quite clear that any realistic calculation of
the problem must take them into consideration.

Recently in~\cite{kn:Pauls95} and \cite{kn:Zank} the simplified approach
\cite{kn:Bar82} was modified by developing the multifluid description for
the neutrals which tried to take into account the highly non-Maxwellian
nature of the neutral distribution. Since in these papers the plasma was
assumed Maxwellian only in each region between the discontinuity surfaces,
the neutral population produced by the charge exchange in these regions has
the same basic characteristics as those of the plasma component.
Although this approach still remains approximate, one must admit that
it allowed its developers to obtain the solutions of the three-dimensional
and nonstationary problems which have not been solved yet on the basis
of the Monte-Carlo method.

\section{Hydrodynamic instabilities in the SW--LISM interaction}

As was admitted in \cite{kn:Wallis76}, the contact discontinuity between the
inner and the outer subsonic region is potentially the subject to the fluid
Kelvin--Helmholtz or MHD flute instabilities which could produce some of
the inhomogeneous structures of astropause tails. The study of
instabilities accompanying the problem of the SW--LISM interaction with
and without magnetic field is an important independent task which lies
outside the scope of this review. We are going to clarify here only the
effect of the choice of numerical methods on their origin and
development in the problem in which  two flows  possessing
substantially different entropies collide at a contact surface calculated
by shock-capturing methods. It is more or less clear that
discontinuity-fitting methods are hardly applicable for this task.

As  was convincingly shown in \cite{kn:Mats92} and \cite{kn:Shima},
complicated nonstationary patterns can be obtained owing to the
Rayleigh--Taylor instabilities
if the interaction is accompanied by the gravitational effect from the star.
In the case of the solar wind the gravity can be neglected, since the
Hoyle--Littleton accretion radius is about 4.5 \AU, which is much smaller
than the expected location of the termination shock. In \cite{kn:Mats89},
however, the stellar wind--interstellar medium  interaction was studied
for parameters different from those corresponding to the solar wind and
the results were obtained clearly indicating instability of the flow. The
Osher numerical scheme \cite{kn:Chak} was used and an algebraically generated
irregular grid rather than the polar one applied in the previous study of
the problem \cite{kn:Shima}. That type of the grid is supposed to give a
finer resolution at large distances from the ejecting center. The
influence of the interstellar medium Mach number was studied in a wide range
from $M_\infty=0.6$ to 15. A highly nonstationary solution was obtained
in this case, instabilities originating both near the astropause
stagnation point and in its lateral region (Fig.~9).

\begin{figure}[t]
\centerline{\psfig{figure=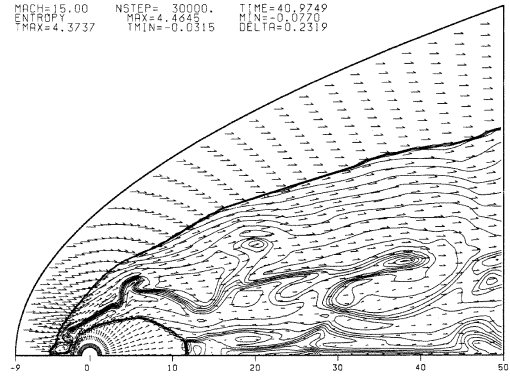,height=5cm}}
\vspace{-6pt}
\caption{ Entropy contours and velocity vectors~[55] ($M_\infty=15$)}
\vspace{-12pt}
\end{figure}

The Kelvin--Helmholtz
instability in the latter region are quite possible as the gases with
different entropy move along the both sides of a contact surface at
different speeds. The origin of the Kelvin--Helmholtz instability near
the stagnation point, however, is questionable, since velocities are very
small there. Such stagnation point instabilities, however, were found earlier
in the calculations of astrophysical jets \cite{kn:Norman82},
\cite{kn:Norman83}, and \cite{kn:Spreiter}. Similar instabilities were
also found in \cite{kn:Mats90a} and \cite{kn:Mats90b} in the calculations
of opposing and forward-facing jets. They were also observed in the
experiments \cite{kn:Finley}, \cite{kn:Romeo63}, and \cite{kn:Romeo65}.
It is worth mentioning, however, that the instabilities of the latter
type were not obtained in the calculations  for parameters close to the
solar wind -- interstellar medium interaction, as was reported in
\cite{kn:Stein94a}. Moreover, N. Pogorelov in his calculations
occasionally obtained instabilities near the stagnation point, even for a
rather rough mesh, using the Roe-type MUSCL scheme under unfavorable
selection of the parameter reconstruction method. Note that one or
another parameter interpolation based on the assumption of the linear,
parabolic \cite{kn:Colella84}, or  higher-order distributions inside
computational cells is usually used to increase the order of accuracy of
the chosen numerical scheme. Say, in \cite{kn:Po93} and
\cite{kn:Po95} the interpolation of characteristic variables was used which
had been discovered to give much more stable results in the vicinity of the
geometrical singularities at $\theta=0$ and $\theta=\pi$. Note that both
the polar grid used in \cite{kn:Po95} and the algebraically generated O-type
grid incorporated into the algorithm \cite{kn:Mats89} possess such a
singularity. In \cite{kn:Po97} the choice of the form of governing equations
(2) or (3) was claimed to have a certain effect on the stability of results.

There also exits another aspect of the problem. As was mentioned in the
previous section, the interaction between the interstellar atoms and the
solar plasma ions via resonant charge-exchange collisions greatly affects
the steady-state structure of the global heliosphere. The flow of
interstellar ions is diverted around the nose of the heliopause, whereas
the neutral particles penetrate into the heliosphere impeded aside from
the charge-exchange collisions.  Thus, there is a velocity difference
between them near the nose which performs like a drag force acting on the
plasma (see Eq.~28). Since the LISM density is larger at the heliopause
than that of the SW, the contact surface is potentially
Rayleigh--Taylor unstable. That kind of instability was admitted in
\cite{kn:Liewer} and \cite{kn:Zank}. Its origin was confirmed by
several numerical experiments, including the one applying the entirely
different particle-in-cell code \cite{kn:Brackbill91}, and by comparing the
instability linear growth rate with the theoretical one. However, as was
admitted earlier, the multifluid hydrodynamic model
in which  all fliuds were supposed to be in a local thermal equilibrium
was used in the above-mentioned papers to determine the motion
of the neutral particles. This makes the obtained results questionable from
the viewpoint of the spatial scale of instabilities. Paper
\cite{kn:Liewer} also neglects the
effect of energetic solar wind neutrals created by the charge exchange
inside the heliopause. The drag force caused by these neutrals tends to
compensate the inward force described earlier and partially suppress the
instability.

The study of hydrodynamic instabilities originating in the SW--LISM
interaction problem is far from its conclusion also due to a number of
physical phenomena that can affect it, such as, the interstellar magnetic
field, cosmic rays, etc. It is important to admit in this review that,
investigating instabilities, one must be very careful in order to
distinguish those of physical and of numerical origin.

\section{Nonstationary SW--LISM interaction}

As was mentioned above and is widely accepted, the solar wind changes its
speed from supersonic to subsonic through a termination shock. A number
of reasons can cause temporal asymmetries of the termination shock and,
therefore,
the interaction pattern as a whole. Among them are disturbances of the
solar wind and its 11-year periodicity. Due to these reasons the
heliospheric shock will move  in response to variation in upstream solar
wind conditions. In \cite{kn:Suess93},  a kinematic analysis  is made of
the solar wind driven temporal variations in the heliospheric
termination shock distance. In \cite{kn:Barnes93}, \cite{kn:Barnes94}, and
\cite{kn:Naidu} the motion of this shock was analyzed
analytically on the basis of one-dimensional gasdynamic model. It was
admitted that the termination shock would rather be nonstationary and
would resemble a distorted asymmetric balloon with some part moving inward
and others moving backward. In~\cite{kn:Stein94b} the termination shock
response to large-scale solar wind fluctuations were studied numerically
using the Lax--Wendroff scheme. The LISM flow was assumed subsonic.
In~\cite{kn:Karmesin}, on the
basis of the similar numerical method 11-year solar wind variation
influence on the inner shock was investigated. In~\cite{kn:Po93}
and~\cite{kn:Po95}
nonstationary problems were modeled by a numerical solution of the Euler
gasdynamic equations~(2) in the finite-volume formulation~(21) using a
MUSCL-type TVD high-resolution numerical scheme.
The suggestion was made of a piecewise-linear distribution  of  the
characteristic parameters inside the cells to determine the values at their
boundaries and slope limiters were used to attain a TVD
property. The results presented below were obtained using the
formulas~\cite{kn:Albada}:
\begin{eqnarray}
  & &  \bar{\bf E}_{l+1/2,n} = \bar{\bf E}({\bf U}^{L},{\bf U}^{R}), \\
  & &  {\bf U}^{L} = {\bf U}_{l,n}+ {\bf U}'_{l,n} \Delta R /2 ,
  \nonumber\\
  & &  {\bf U}^{R} = {\bf U}_{l+1,n}- {\bf U}'_{l+1,n} \Delta R /2 ,
  \nonumber\\
  & &  {\bf U}'_{l,n} = S_{l,n} {\bf W}'_{l,n} ,\nonumber\\
  & &  {\bf W}'_{l,n} = {( b^{2}_{m}+c ) a_{m} + ( a^{2}_{m}+c) b_{m}\over
      a^{2}_{m} + b^{2}_{m} + 2c} ,\nonumber \\
  & &  {\bf a} = S^{-1}_{l,n} ({\bf U}_{l+1,n}- {\bf U}_{l,n}),
  \nonumber \\
  & &  {\bf b} = S^{-1}_{l,n} ({\bf U}_{l,n}- {\bf U}_{l-1,n})\nonumber
\end{eqnarray}

Here $c$ is a small positive value used to avoid division by  zero.  In
these formulas $S$ and~$S^{-1}$ are $(4\times 4)$ matrices,
constructed using  right and left eigenvectors of the Jacobian matrix
$\partial\bar{\bf E}/ \partial {\bf U}$.

The fluxes presented by Eq.~(30) were defined  on  the  basis  of  Roe's
approximate Riemann solver~\cite{kn:Roe81}.

Fluxes  through  another  pair  of  cell  surfaces  can be   obtained
similarly.

The promotion of the solution in time was  performed  in  the  following way:
\begin{eqnarray*}
  & &{\bf U}^{(1)}_{l,n} = {\bf U}^{k}_{l,n}+ {\Delta t\over 2}\,
      {\partial {\bf U}^k_{l,n}\over \partial t};\\
  & &{\bf U}^{k+1}_{l,n} = {\bf U}^{k}_{l,n}+ \Delta t\,
    {\partial {\bf U}^{(1)}_{l,n}\over \partial t} ,\\
\end{eqnarray*}

\noindent
where $t = k\, \Delta t$, $k=0,1,\ldots\,$, and $\Delta t$ is defined by
the time resolution and by the CFL condition.

The LISM proton  number density was assumed to be
$n_{H+} = \rm 1\, cm^{-3}$.  The velocity of LISM relative to  the
solar system is about $\rm20\, km/s$, while the speed of sound of  the  LISM
gas is about $\rm10\,km/s$.
The SW protons number density was  adopted
$n_{H+}= \rm10\, cm^{-3}$, while its velocity is
$V_{e} \approx \rm500\, km/s$, the speed of sound $c_{e}\approx \rm100\, km/s$
at the distance of  the Earth's orbit ($1\,\AU$).
This corresponds to the following values of dimensionless parameters
chosen for
the initial data: $M_{\infty }=2$, $M_{e}=5$, $\chi  =400$, $K = 6250$.
The stationary initial flow was obtained by the
same numerical method using a time-stabilization approach.  The
calculation was performed in the
ring  region with the inner and outer circle radii being $R_{\rm min }= 10$
and $R_{\rm max }= 500\,\AU$.
On the inner surface all parameters were specified  as  functions
of time by formulas:
\begin{eqnarray*}
  & &U= 25,\quad \rho (t) = \rho (0)(1 + 4\exp [-(t-3)^{2}]),
  \\
  & &p(t) = 3.2325 \rho ^{\gamma }(t),
\end{eqnarray*}

\noindent
since this boundary is supersonic. Here~$U$ is  the  radial  velocity
component. Although the choice of parameters is somewhat nonrealistic, as
far as the solar wind is concerned, it still can be used for a
qualitative analysis of a nonstationary picture of the interaction.

The initial distributions of pressure (below the symmetry axis) and density
logarithms are shown in Fig.~5. The size of the outer circle in the
figures is 400 \AU. Similar   isolines are presented in Figs.~\hbox{10--13}
at different moments of time (in the  units $1\,\AU/U_{\infty }$).  The
growing part of the  disturbance,  interacting  with the inner shock, moves
it, first,  from  the  centre,  Fig.~10,  the intensity of~IS increasing.
Later, it tends to  achieve  the  initial position.

\begin{figure}[t]
\centerline{\psfig{figure=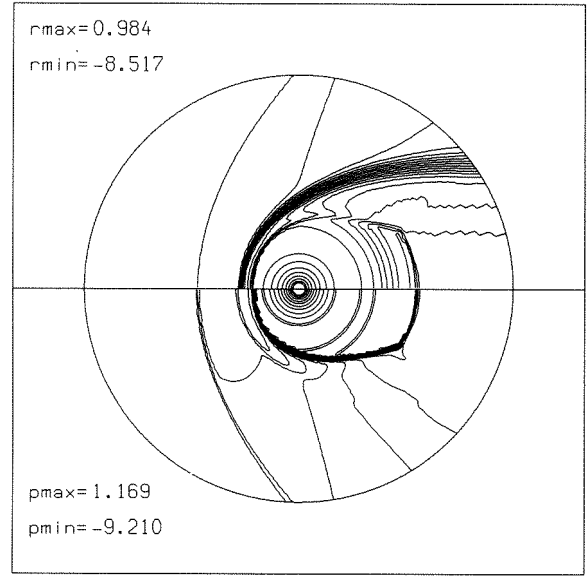,height=7.7cm}}
\vspace{-6pt}
\caption{
Pressure and density logarithm isolines [75], $t=8$
}
\vspace{-12pt}
\end{figure}

\begin{figure}[t]
\centerline{\psfig{figure=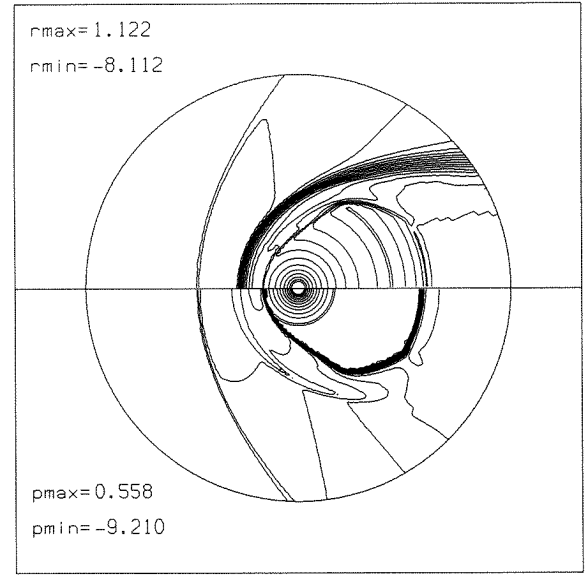,height=7.7cm}}
\vspace{-6pt}
\caption{
Pressure and density logarithm isolines [75], $t=12$
}
\vspace{-12pt}
\end{figure}

The penetration of the disturbance in the region between IS and~CD
is seen on both charts of isolines at different moments of time.  The
disturbance, while crossing~IS, increases  greatly,  and  after  some
time a local pressure  maximum originates between IS  and~CD (Fig.~11).
This  leads  to  the
effect of suction of SW gas to the centre accompanying IS  motion  in
the same direction.
This pressure  extremum  line  gradually  moves
towards~CD, see Figs.~\hbox{12--13}.\looseness=1

\begin{figure}[t]
\centerline{\psfig{figure=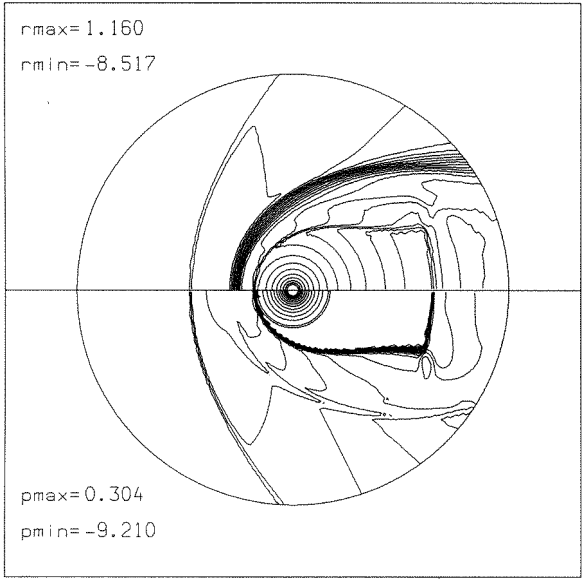,height=7.7cm}}
\vspace{-6pt}
\caption{
Pressure and density logarithm isolines [75], $t=18$
}
\vspace{-12pt}
\end{figure}

\begin{figure}[t]
\centerline{\psfig{figure=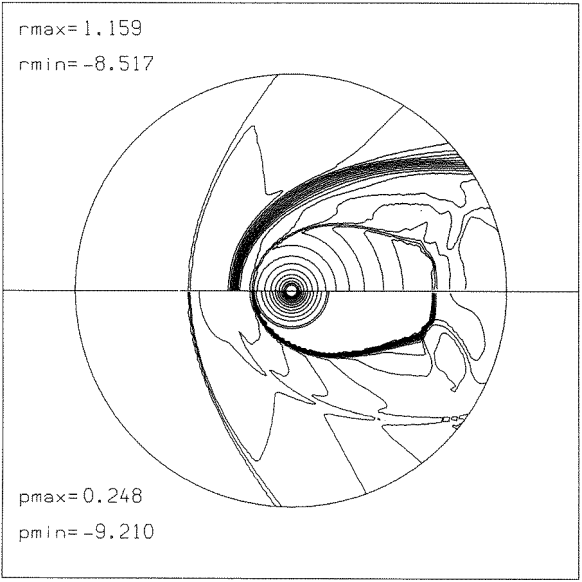,height=7.7cm}}
\vspace{-6pt}
\caption{
Pressure and density logarithm isolines [75], $t=21$
}
\vspace{-12pt}
\end{figure}

The parts of the computational region more remote from the  source
position suffer the same changes later in time.

As soon as the disturbance  meets  different
points  of~IS,  the
latter suffers substantial distortion. When the  triple point  on~IS  is
reached, there appear  two  triple  points  with  different  reflected
shocks and recirculation zone between them (Fig.~13). As  soon  as  the
source intensity becomes constant, the position of IS gradually,  but
rather  slowly,  moves  towards  its  initial  position.  The  maximum
increase of MD stand-off distance is greater than that of~IS  along  the
ray $\theta  = 0$.\looseness=1

The relaxation of the flow towards  some  stationary  solution  is
very  slow ($t>220$). When $t>20$,  in   the   region   near $\theta =0$
additional local pressure extrema vanish. The  extended  vortex  region
originates behind MD and it is up  to $t = 60$,  when  these  vortices
vanish, probably due to numerical viscosity.

Now we present some results~\cite{kn:Po95} concerning the periodic SW--LISM
interaction. In this calculation the LISM proton number density, its
velocity and the speed of sound are $n_{\infty}= \rm 0.1\, cm^{-3}$,
$V_\infty=\rm 20\, km/s$ and $c_\infty=\rm 10\, km/s$, respectively.
Parameters of the  solar wind change substantially within the 11-year
period of the solar activity.
The concentration of charged particles $n_e=\rm 1.56\, cm^{-3}$ and the
 radial SW velocity $V_e=\rm 400\, km/s$ are chosen for the minimum of
the solar activity, while $n_e=\rm 8\, cm^{-3}$ and $V_e=\rm 500\, km/s$
correspond to its maximum (see \cite{kn:Bar77} and \cite{kn:Brandt}).
Thus, for dimensionless
parameters we have $K=6250$, $\chi=256$, $M_\infty=2$, and $M_e=5$ in the
minimum and  $K=50000$, $\chi=400$, $M_\infty=2$, and $M_e=5$ in the
maximum of the solar activity. These values are chosen as the basic
points for the sine function approximating the time dependence of $K$ and
$\chi$ within 11 years. The dimensionless time unit is 86.8 days.
The time step is chosen to be $\Delta t \approx 0.4$ days. Parameter
distribution corresponding to the stationary solution in the minimum
solar activity is chosen as initial data. The results are obtained by the
MUSCL numerical method  described above. The calculation is
performed
in the ring region with the inner and the outer circle radii being
$R_{\rm min} =14$ and $R_{\rm max}=700$, respectively.

The isolines of the pressure (below the symmetry axis) and the density
logarithms corresponding to the initial data are shown in Fig.~14.
They are presented for the outer circle size equal to 560 \AU. The number
of cells is 99 and 116 in the radial and in the angular direction,
respectively. At the initial stage of the flow development, the increase
of the parameter $K$ results in the TS motion from the Sun. The motion of
the compression wave through TS causes the origin of a new shock wave
propagating  from the center. This can be seen in Fig.~15 corresponding
to $t=20$. This shock penetrates through the contact discontinuity and
moves towards the bow shock, see Fig.~16 ($t=36$).

\begin{figure}[t]
\centerline{\psfig{figure=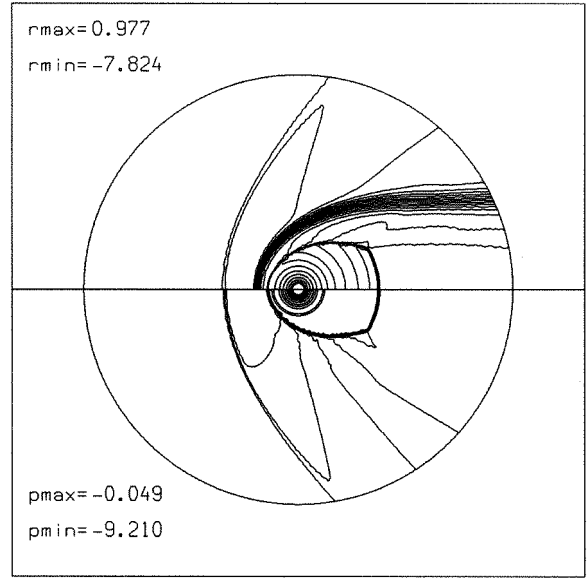,height=7.7cm}}
\vspace{-6pt}
\caption{
Pressure and density logarithm isolines [76], $t=0$
}
\vspace{-12pt}
\end{figure}

\begin{figure}[t]
\centerline{\psfig{figure=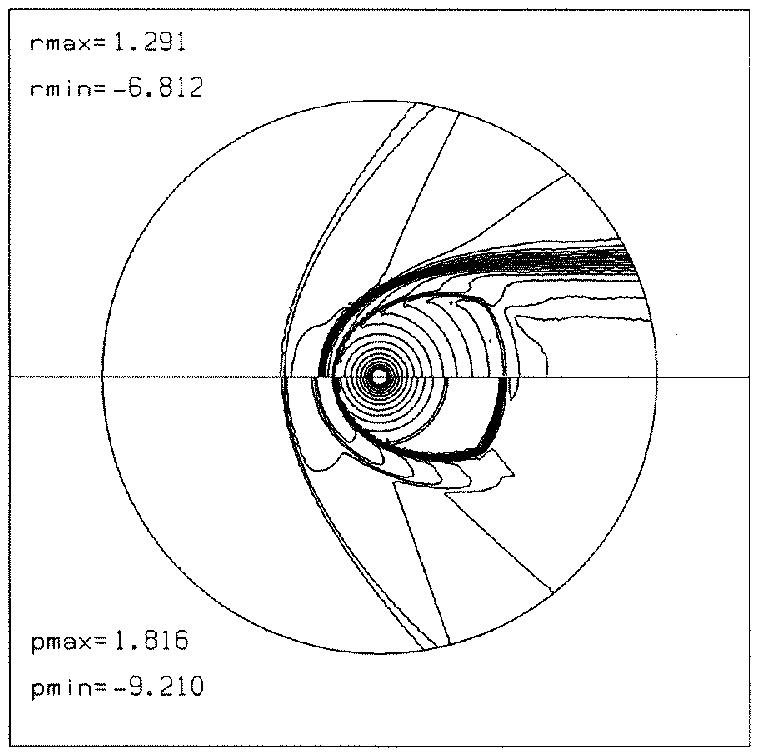,height=7.7cm}}
\vspace{-6pt}
\caption{
Pressure and density logarithm isolines [76], $t=20$
}
\vspace{-12pt}
\end{figure}

\begin{figure}[t]
\centerline{\psfig{figure=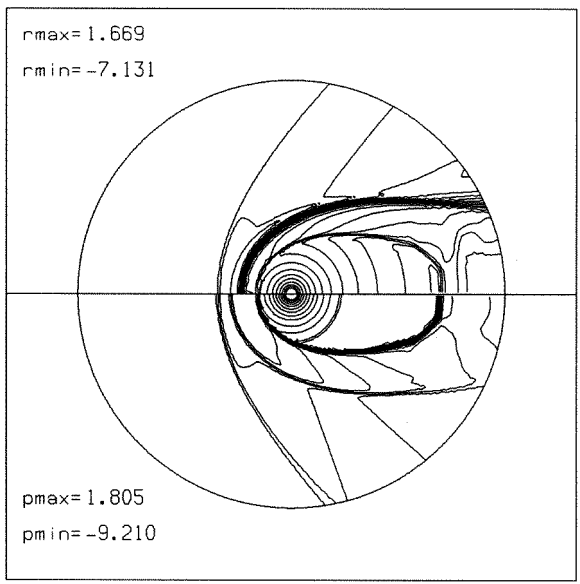,height=7.7cm}}
\vspace{-6pt}
\caption{
Pressure and density logarithm isolines [76], $t=36$
}
\vspace{-12pt}
\end{figure}

The decreasing part
of the periodic function causes a backward motion of the termination
shock towards the Sun. This leads to the origin of new flow division
surfaces, reverse flow zones, and vortices  of variable size and intensity
in the wake region. At $t=60$, see Fig.~17, the size of the bullet shape
termination shock becomes minimum again. Later on a next nonstationary
shock wave appears, etc. A definite 11-year periodicity is developed in
the shape of the termination shock. Parameter distribution between the
inner and the bow shock is determined by the propagation of
shock waves traveling one after another and interacting
with the less intensive waves reflected from the bow shock (Fig.~18).

\begin{figure}[t]
\centerline{\psfig{figure=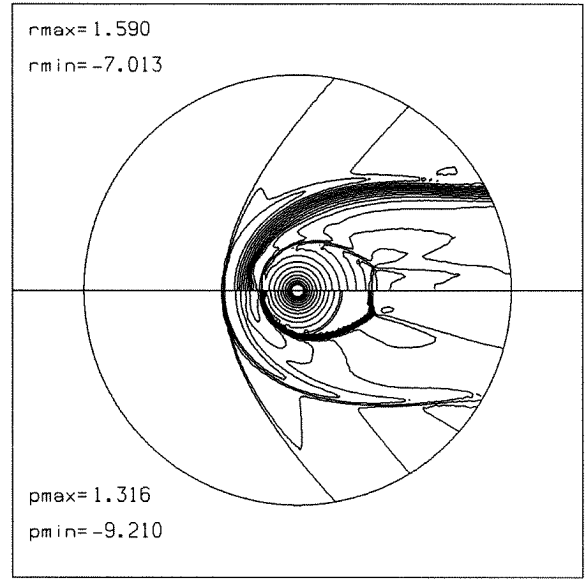,height=7.7cm}}
\vspace{-6pt}
\caption{
Pressure and density logarithm isolines [76], $t=60$
}
\vspace{-12pt}
\end{figure}

\begin{figure}[t]
\centerline{\psfig{figure=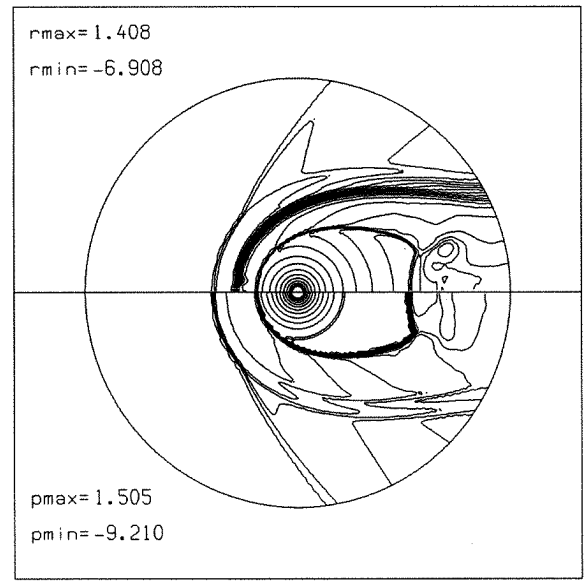,height=7.7cm}}
\vspace{-6pt}
\caption{
Pressure and density logarithm isolines [76], $t=84$
}
\vspace{-12pt}
\end{figure}

This modeling shows that the flow pattern is
substantially nonstationary which must be taken into account when
analyzing responses from the space vehicles crossing these discontinuities.
Variable solar activity is usually accompanied by the asymmetry of the solar
wind which will be considered in the next section.

Application of a high-resolution numerical methods allows one to avoid
spurious oscillations near shocks and provides their sharper resolution
using smaller number of computational cells than it is necessary in
nonmonotone schemes with artificial viscosity.

\section{Nonuniform solar wind -- interstellar medium interaction}

Distant solar wind deviations from spherical symmetry induced by the
interaction with neutral interstellar hydrogen due to photoionization and
charge-exchange processes were studied in \cite{kn:Grz75} by a
perturbation technique. The penetration of neutral particles deep inside
the heliosphere results in a substantial increase of the distant solar
wind temperature. In all models described in the previous sections we
considered the stellar (solar) wind as spherically-symmetric. It can
fairly easily become asymmetric, since charge-exchange processes are
clearly more effective in the forward region of the interaction because their
efficiency is proportional to the velocity difference between neutral
and charged particles. The perturbation analysis has shown that a
strongly asymmetric distribution of the neutral hydrogen within the
heliosphere causes asymmetric deceleration and extremely nonuniform
distribution of the distant solar wind temperature, thus leading to
non-radial gradients and flows.

This is, however, only one (external) reason of the solar wind asymmetry.
Another one is determined by a mere variation of the solar wind symmetry
in time within its 11-year activity period (see, e.g., speculations in
\cite{kn:Po95}). According to the solar minimum observations by Ulysses,
the solar wind parameters depend on the helioaltitude. These data (see
\cite{kn:Phillips95} and \cite{kn:Phillips96}) indicate that two large
polar coronal holes, one in the northern and the other in the southern
hemisphere, produce a hotter, lower-density, higher-speed wind comparing
with the ecliptic wind. In \cite{kn:Pauls96} the calculations were
performed of a nonuniform solar wind--interstellar medium interaction
using the earlier mentioned three-dimensional time-dependent ZEUS method
\cite{kn:Stone}. The solar wind
boundary conditions were taken from the published Ulysses data. According
to them, the ram pressure increases 1.5 times from the ecliptic plane to
the solar pole. That is why, the termination shock in the calculations
was found to be elongated along the solar axis which, in turn, resulted
in an increased flow in the ecliptic plane compared with that over the
solar poles. The authors reported a  pronounced effect of the solar wind
asymmetry on the global structure of the termination shock and heliopause.
Both a two-shock (supersonic LISM) and a one-shock (subsonic LISM) model were
considered. It is worth mentioning once again in this connection that, if
only the motion of charged particles is considered, the LISM flow is
definitely supersonic. Pure gasdynamic subsonic models for the problem
under consideration are sometimes used to account for the net effect of
the charge-exchange  processes, the influence of cosmic rays
\cite{kn:Fahr92}, and interstellar magnetic field. The latter subject
will be discussed in the next section.

The paper \cite{kn:Pauls96} definitely indicates, from our viewpoint,
that any realistic model for the SW--LISM interaction must include both
neutral particles, the asymmetry of the solar wind, and, as a consequence,
the solar wind periodicity, since the asymmetry varies within the solar
activity period. This, of course, does not prevent an investigation of
different physical effects separately.

\section{Solar wind interaction with the magnetized interstellar medium}

The presence of the interstellar magnetic field necessitates solution
of the MHD equations for modeling of the SW--LISM interaction. The
influence of magnetic field becomes important if a magnetic pressure
${\bf B}^2/8\pi$ becomes comparable with a dynamic  pressure. Magnetic
field causes an increase of the maximum speed of small perturbations in
the LISM flow, especially in the direction normal to the direction of the
magnetic field, thus resulting in the decrease of the effective Mach
number \cite{kn:Fahr86} and \cite{kn:Fahr88}. Though the magnitude and
the direction of the interstellar magnetic field is not perfectly known,
the estimates from \cite{kn:Axford} indicate the possible importance of
its presence. The value and the direction of the LISM magnetic field can
affect the global structure of the interaction not only directly, but
also by modulating the distance between the bow shock and the heliopause
which determines the transparency of this layer for the LISM neutrals.
This, in turn, affects the observed line profiles of the solar
Lyman-$\alpha$ backscattered emission and estimates of the LISM parameters
(see the discussion in \cite{kn:Bar95} and \cite{kn:Lal}).
If the LISM magnetic field is not parallel to its velocity, the problem
becomes three-dimensional. Later in the section we discuss the results of
such MHD modeling.

It is worth mentioning, that although the value of the
solar magnetic field is often neglected in numerical modeling of the
problem under consideration at the distances of the termination shock,
its influence is substantial at Earth's magnetosphere distances
\cite{kn:Ogino} and \cite{kn:Watanabe}. In \cite{kn:Washimi} the global
structure of the outer heliosphere was studied in the axisymmetric
formulation for a subsonic interstellar medium with taking into account
a time-varying poloidal magnetic field.
In \cite{kn:Washimi96} and \cite{kn:Nozawa} the toroidal magnetic field
in the heliosheath was found to increase with the distance from the Sun.
The complicated three-dimensional nostationary behavior of the flow was
studied which showed the importance of taking this component of the
magnetic field into account.  We restrict ourselves to noting this aspect
of the problem and will not discuss it below.

\subsection{On the eigenvector and eigenvalue systems of MHD equations}

The system of ideal MHD equations in the conservation-law form is presented
by Eq.~(1). This system can be directly rewritten in the quasi-linear form
which is more convenient for the characteristic analysis. It is
well-known that the equation $\div {\bf B} =0$
expresses the absence of magnetic charge. It is also evident that, if
magnetic charge is absent initially, it will not appear
mathematically at any time instant. From this viewpoint the above
equation is excessive. If we rewrite the system of MHD
equations~\cite{kn:Landau}
\begin{eqnarray*}
& &{\partial \rho \over \partial t} + \div \rho{\bf v} =0, \\
& &{\partial {\bf v} \over \partial t} + ({\bf v}\cdot\nabla){\bf v}=
-{\nabla p \over \rho} - {{\bf B}\times \rot {\bf B} \over 4\pi\rho}, \\
& & {dp \over dt} = a {d\rho \over dt} \\
& & \pd{{\bf B}}{t} = \rot ({\bf v}\times {\bf B})
\end{eqnarray*}

\noindent
in quasilinear form  not paying attention to the condition $\div {\bf B}=0$, the
following system is obtained ($a$ is the acoustic speed of sound):
\begin{equation}
  {\partial {\bf u}\over \partial t} +
  A_1\,{\partial {\bf u}\over \partial x} +
  A_2\,{\partial {\bf u}\over \partial y} +
  A_3\,{\partial {\bf u}\over \partial z}  = \mathchar"0630\, ,
\end{equation}

\noindent
where
\[
{\bf u}  =  \left(\rho,\, u,\, v,\, w,\, p,\, B_x,\, B_y,\,
B_z\right)^T,
\]

\[
\medmuskip=0mu
A_1=\left(
\begin{array}{cccccccc}
u & \rho & 0 & 0 & 0 & 0 & 0 & 0 \\
0 & u & 0 & 0 & {1\over\rho} & 0 & {B_y\over 4\pi\rho} &
{B_z\over 4\pi\rho} \\
0 & 0 & u & 0 & 0 & 0 & -{B_x\over 4\pi\rho} & 0 \\
0 & 0 & 0 & u & 0 & 0 & 0 & -\frac{B_x}{4 \pi\rho} \\
0 & \rho a^2 & 0 & 0 & u & 0 & 0 & 0 \\
0 & 0 & 0 & 0 & 0 & u & 0 & 0 \\
0 & B_y & -B_x & 0 & 0 & 0 & u & 0 \\
0 & B_z & 0 & -B_x & 0 & 0 & 0 & u
\end{array}
\right)\!,
\]

\[
\medmuskip=0mu
A_2=\left(
\begin{array}{cccccccc}
v & 0 & \rho & 0 & 0 & 0 & 0 & 0 \\
0 & v & 0 & 0 & 0 & -{B_y\over 4\pi\rho}  & 0 & 0 \\
0 & 0 & v & 0 & \frac{1}{\rho} & {B_x\over 4\pi\rho}& 0 &{B_z\over 4\pi\rho}\\
0 & 0 & 0 & v & 0 & 0 & 0 & -\frac{B_y}{4 \pi\rho} \\
0 & 0 & \rho a^2 & 0 & v & 0 & 0 & 0 \\
0 & -B_y & B_x & 0 & 0 & v & 0 & 0 \\
0 & 0 & 0 & 0 & 0 & 0 & v & 0 \\
0 & 0 & B_z & -B_y & 0 & 0 & 0 & v
\end{array}
\right)\!,
\]

\[
\medmuskip=0mu
A_3=\left(
\begin{array}{cccccccc}
w & 0 & 0 & \rho & 0 & 0 & 0 & 0 \\
0 & w & 0 & 0 & 0 & -{B_z\over 4\pi\rho}  & 0 & 0 \\
0 & 0 & w & 0 & 0 & 0 &-{B_z\over 4\pi\rho}& 0 \\
0 & 0 & 0 & w & \frac{1}{\rho} & \frac{B_x}{4 \pi\rho}& \frac{B_y}{4
\pi\rho} & 0  \\
0 & 0 & 0 & \rho a^2 & w & 0 & 0 & 0 \\
0 & -B_z & 0 & B_x & 0 & w & 0 & 0 \\
0 & 0 & -B_z & B_y & 0 & 0 & w & 0 \\
0 & 0 & 0 & 0 & 0 & 0 & 0 & w
\end{array}
\right)\!,
\]

Solution of the characteristic equation
\[
\det(A_1-\lambda I) =0
\]

\noindent
gives the following eigenvalues:
\begin{eqnarray}
& &\lambda_{1,2}=u, \quad \lambda_{3,4}=u \pm \frac{B_x}{\sqrt
{4\pi\rho}},\\
& &\lambda_{5,6,7,8}=u\pm{1\over 2}\biggl[
\biggl(a^2+\frac{{\bf B}^2}{4\pi\rho}+
\frac{B_x a}{\sqrt{\pi\rho}}\biggr)^{1/2} \nonumber
\\
& &\pm \biggl(a^2+\frac{{\bf B}^2}{4\pi\rho}-
\frac{B_x a}{\sqrt{\pi\rho}}\biggr)^{1/2}\biggr]
\end{eqnarray}

Note that in one-dimensional treatment the equation for $B_x$ reduces
to
\begin{equation}
{\partial B_x \over \partial t} + u{\partial B_x\over \partial x}=0,
\end{equation}

\noindent
that is, to the one-dimensional convection equation for $B_x$.
Of course, in the truly one-dimensional problem (all values depend only on
the spatial variable $x$) one can simply assume $B_x \equiv \const$
and omit the corresponding equation. On the contrary, if we are going to
apply the solution of the one-dimensional MHD Riemann problem to determine
the flux through the cell boundary, such an assumption is too excessive,
since only the integral $\oint B_n d \sigma$ over the whole
computational cell must be equal to zero. Among  eigenvalues (32)--(33)
the first two correspond to the entropy and $B_x$ convection waves,
$\lambda_{3,4}$ corresponds to the Alfv\'en, or rotational, waves, and
the other ones to the slow and to the fast magnetosonic wave. Omitting
Eq.~(34), we reduce the system to $7\times 7$. Both the extended
$8\times 8$ system and the reduced one have real eigenvalues and a
degenerate set of eigenvectors. One can easily derive expressions for
them. Otherwise, one can refer to \cite{kn:Brio}, \cite{kn:Powell94}, and
\cite{kn:Sun}. As was admitted in \cite{kn:Gombosi} and
\cite{kn:Powell95}, we can use the
extended system to derive an approximate solution to the MHD Riemann problem.
Another possibility is to derive formulas for the $7\times 7$ system and use
a convection equation to find $B_x$ ($x$ is normal to the cell boundary)
on the cell surface. It is also useful to realize that by collecting the
source term from system~(31) to arrive at the conservative form similar
to~(1), we obtain the following system:
\begin{equation}
  {\partial {\bf U}\over \partial t} +
  {\partial {\bf E}\over \partial x} +
  {\partial {\bf F}\over \partial y} +
  {\partial {\bf G}\over \partial z} + {\bf H_{\rm div}} = \mathchar"0630\, ,
\end{equation}

\noindent
where
\[
  {\bf H_{\rm div}}= \div {\bf B} \left(
  0,\,{B_x \over 4\pi},\,{B_y \over 4\pi},\,{B_z \over 4\pi},\,
  {{\bf v}\cdot {\bf B}\over 4\pi},\,u,\,v,\,w\right)^{\rm T}
\]

This form of the system will be used later to satisfy the divergence-free
condition.

Since the application of high-resolution numerical schemes to MHD flows
and to the problem under consideration, in particular, has not yet become
common~(see some tests in \cite{kn:Toth}), we give their
description in the next subsection.

\subsection{High-resolution numerical schemes for MHD equation}

TVD upwind and symmetric differencing schemes have recently become very
efficient tool for solving complex multi-shocked gasdynamic flows. This is
due to their robustness for strong shock wave calculations. A general
discussion of the modern high-resolution shock-capturing methods and their
application for a variety of gasdynamic problems can be found in
\cite{kn:Hirsch} and \cite{kn:Yee}.  The extension of these schemes to the
equations of the ideal magnetohydrodynamics (MHD) is not straightforward.
First, the exact solution  \cite{kn:Kulik65} of the MHD Riemann problem is
too multivariant to be used in regular calculations.  Second, several
different approximate solvers \cite{kn:Brio}, \cite{kn:Cargo},
\cite{kn:Dai94a}, \cite{kn:Hanawa}, \cite{kn:Po95t}, \cite{kn:Powell95},
and \cite{kn:Zachary} applied to MHD equations are now at the stage of
investigation and comparison.

The schemes
\cite{kn:Brio}, \cite{kn:Cargo}, \cite{kn:Hanawa},
and~\cite{kn:Po95t}, \cite{kn:Powell95}
are based on the MHD extensions of Roe's linearization
procedure \cite{kn:Roe81}. In \cite{kn:Brio}, the attempt of such extension
was made and the second order upwind scheme was constructed that
demonstrated several advantages in comparison with the Lax--Friedrichs,
the Lax--Wendroff, and the flux-corrected transport scheme~\cite{kn:Devore}.
Roe's procedure, however, turned out to be realizable only for the special
case with the specific heat ratio $\gamma = 2$. The reason of such behavior
of MHD equations is that there is not any single averaging procedure to
find a frozen Jacobian matrix of the system.  Another linearization
approach is used in \cite{kn:Cargo}, \cite{kn:Hanawa},
\cite{kn:Po95t}, and \cite{kn:Powell95}
in which  the linearized Jacobian matrix is not a function of a
single averaged set of variables, but depends in a complicated way on the
variables on the right- and on the left-hand side of the computational cell
surface. In \cite{kn:Po96p} and \cite{kn:Po97zh}  this procedure was shown
to be nonunique.  A multiparametric family of linearized MHD approximate
Riemann problem solutions was presented that assured an exact satisfaction
of the conservation relations on discontinuities. A proper choice of
parameters is necessary to avoid physically inconsistent solutions.

Consider the one-dimensional system of MHD equations

\begin{equation}
{\partial {\bf U} \over \partial t} + {\partial {\bf F} \over \partial x} =
\mathchar"0630,
\label{thesystem}
\end{equation}
where ${\bf U}  =  \left(\rho,\, \rho u,\, \rho  v,\, \rho w,\, e,\, B_y,
\,B_z\right)^{\rm T}$ and

\[
{\bf F}({\bf U}) = \left( \begin{array}{c}
\rho u \\ \rho u^2+p_0-B_x^2/4 \pi \\ \rho u v - B_x B_y /4 \pi \\
\rho u w -B_x B_z/4 \pi \\ (e+p_0)u-(u B_x+v B_y+w B_z) B_x/4 \pi \\
u B_y-v B_x \\ u B_z-w B_x
\end{array}
\right)
\]

In these formulas $e=p/(\gamma -1)+\rho (u^2+v^2+w^2)/2+
(B_x^2+B_y^2+B_z^2)/8 \pi$ is the total energy per unit volume,
$p_0=p+(B_x^2+B_y^2+B_z^2)/8 \pi$ is the total pressure, $p$ and $\rho$ are
pressure and density, ${\bf v}=(u,v,w)$ is the velocity vector, ${\bf
B}=(B_x,B_y,B_z)$ is the magnetic field vector, and $\gamma$ is the adiabatic
index. We assume all functions to depend only on time $t$ and on the linear
coordinate $x$. Our aim is to construct a solution to Eq.~(36) for $t>0$ for
the piecewise-constant initial distribution of $\bf U$: ${\bf U}={\bf U}_1$
for $x<0$ and ${\bf U}={\bf U}_2$ for $x>0$. It is assumed, owing to the
divergence-free condition, that $B_x=B_{x1}=B_{x2} \equiv {\rm const}$.

Let us first find the exact expression for the matrix $A=\Delta {\bf F}/
\Delta {\bf U}$, where $\Delta {\bf U}={\bf U}_1 -{\bf U}_2$ and
$\Delta {\bf F}={\bf F}_1 -{\bf F}_2$. The function $\bf F$ being nonlinear,
the expression for $A$ is determined nonuniquely. In fact, if we choose some
nondegenerate substitution ${\bf s}={\bf s}({\bf U})$, then from the exact
equalities $\Delta {\bf F} = A_F \Delta {\bf s}$ and $\Delta {\bf U} = A_U
\Delta {\bf s}$ it follows that $A=A_F (A_U)^{-1}$. The exact analytic
expressions for $A_F$ and $A_U$ can be written out explicitly if $\bf U$ and
$\bf F$ are fractional linear functions of $\bf s$ or  polynomials with
respect to its components. We use here the equivalent transforms of
the type $\Delta (BC)= \frac{1}{2} (B_1+B_2) \Delta{}C+ \frac{1}{2}
(C_1+C_2) \Delta{}B$.  The structure and the simplicity of $A$ depends on
the choice of $\bf s$.  The matrix $A$ is an approximation to the Jacobian
matrix $\displaystyle J=\frac{\partial {\bf F}}{\partial {\bf U}}$ and must
conserve the main hyperbolic properties of $J$. It must be representable in
the form $A=\Omega_R \Lambda \Omega_L$, where $\Omega_L$ and $\Omega_R$ are
the matrices of its left and right eigenvectors, respectively, $\Omega_R
\Omega_L=I$, and $I$ is the identity matrix; $\Lambda= ||\lambda_i
\delta_{ij}||$ is the  diagonal matrix of the real eigenvalues of $A$, and
$\delta_{ij}$ is the Kronecker delta.

Then the sought solution to Eq.~(36) for $t>0$ acquires the form
\begin{eqnarray}
{\bf U}(\xi) &=& \frac{1}{2} ({\bf U}_1+{\bf U}_2 + \Omega_R S(\xi)
\Omega_L \Delta {\bf U}),\\
{\bf F}(\xi) &=& \frac{1}{2} ({\bf F}_1+{\bf F}_2 + \Omega_R
|\Lambda(\xi)| \Omega_L \Delta {\bf U}) \end{eqnarray}

\noindent
where $\xi = \frac{x}{t}$, $S(\xi)=||\sgn (\lambda_i -\xi) \delta_{ij}||$, and $|\Lambda
(\xi)|=||\lambda_i \sgn (\lambda_i -\xi) \delta_{ij}||$.  Equations~(37)
and~(38) determine the piecewise-constant functions ${\bf U}(\xi)$ and ${\bf
F}(\xi)$ that glue the right and the left values of the initial
distributions via the system of jumps. If the Hugoniot-type condition is
valid $\Delta {\bf F}=\lambda \Delta {\bf U}$, where $\lambda$ is the jump
velocity, then $\lambda$ is one of the eigenvalues of $A$, since $\Delta
{\bf F}-\lambda \Delta {\bf U}=(A_F-\lambda A_U) \Delta {\bf s}=(A-\lambda
E) A_U \Delta {\bf s}=(A-\lambda E) \Delta {\bf U}=0$. Thus, $\det
(A-\lambda E)=0$, $\Delta {\bf U} \neq 0$ is the eigenvector of $A$, and
relations~(37)--(38) describe the jump exactly.

The solution of this kind was first constructed for the equations of ideal
gas dynamics \cite{kn:Roe81}. In \cite{kn:Brio} such solution was given for
Eq.~(36) in the case $\gamma =2$. The approximate solution for an arbitrary
adiabatic index was proposed in \cite{kn:Hanawa}. Here we present the
procedure for obtaining the extension of Roe's linearization procedure for
MHD equation and show that it is not unique. Thus, the solution
from~\cite{kn:Hanawa} is the particular case of the  multiparametric
family of approximate solutions to the MHD Riemann problem.

We choose the vector $\bf s$ as a generalization of that for pure gas
dynamics:

\[
{\bf s}=\left(
\begin{array}{c}
R \\ U \\ V \\ W \\ {\cal I} \\ Y \\ Z
\end{array}
\right)=\left(
\begin{array}{c}
\sqrt {\rho} \\ \sqrt {\rho} u \\ \sqrt {\rho} v \\ \sqrt {\rho} w \\
\sqrt {\rho} H \\ B_y/\sqrt{\rho} \\ B_z/\sqrt {\rho}
\end{array}
\right)
\]

Then

\[
{\bf U}= \left(
\begin{array}{c}
R^2 \\ RU \\ RV \\ RW \\ U_5 \\
RY \\ RZ
\end{array}
\right),
\]

\noindent
where

\begin{eqnarray*}
U_5 &=& \frac{R{\cal I}}{\gamma} + \frac{(U^2+V^2+W^2)(\gamma -1)}{2
\gamma} \\
    &-& \frac{(Y^2+\nobreak Z^2) R^2 (2-\gamma)}{8\pi\gamma}
\end{eqnarray*}

\noindent
and $H=(e+p_0)/\rho$ is the total enthalpy. The vector $\bf F$ in the
variables$\bf s$ acquires the form

\[
{\bf F}= \left(
\begin{array}{c}
RU \\
F_2\\ UV-RYB_x/4 \pi  \\ UW -RZB_x/4 \pi \\
U{\cal I}- (UB_x/R+VY+WZ)B_x/4 \pi \\
UY-VB_x/R \\ UZ-WB_x/R
\end{array}
\right),
\]

\noindent
where $F_2=U^2+(\gamma -1)R{\cal I}/\gamma -(U^2+V^2+W^2)\*
\allowbreak (\gamma -1)/2 \gamma +
(Y^2+Z^2)R^2(2-\gamma)/8 \pi \gamma$.

Using the new expressions for $\bf U$ and $\bf F$, we can find the matrices
$A_U$, $A_F$, and $A_{\lambda}=A_F-\lambda A_U$. We present only the
expression for $A_\lambda$:

\[
\arraycolsep=1pt
\medmuskip=0mu
A_\lambda=\left(
\begin{array}{ccccccc}
u-2\lambda & 1 & 0 & 0 & 0 & 0 & 0 \\
A_{21} & \frac{\gamma +1}{\gamma}u-\lambda & \frac{1-\gamma}{\gamma}v
& \frac{1-\gamma}{\gamma}w & \frac{\gamma -1}{\gamma} & q_y & q_z \\
A_{31} & v & u-\lambda & 0 & 0 & -\frac{B_x}{4 \pi} & 0 \\
A_{41} & w & 0 & u-\lambda & 0 & 0 & -\frac{B_x}{4 \pi} \\
A_{51} & A_{52} & A_{53} & A_{54} &u-\frac{\lambda}{\gamma} &
A_{56} & A_{57} \\
A_{61} & h_y & -\frac{B_x}{\rho} & 0 & 0 & u-\lambda & 0 \\
A_{71} & h_z & 0 & -\frac{B_x}{\rho} & 0 & 0 & u-\lambda
\end{array}
\right)\!,
\]

\noindent where

\begin{eqnarray*}
A_{21} &=& \frac{\gamma -1}{\gamma} {\cal H} -\lambda u+q,
\ A_{31} = -\frac{B_x
h_y}{4 \pi} -\lambda v,   \\
A_{41} &=& \frac{B_x h_z}{4 \pi} -\lambda w,\
A_{51} = \frac{uB_x^2}{4 \pi\rho}+\lambda
\left(q-\frac{{\cal H}}{\gamma}\right), \\
A_{52} &=& {\cal H}-\frac{B_x^2}{4\pi\rho}+\lambda u
\frac{1-\gamma}{\gamma}, \\
A_{53} &=&-\frac{B_x h_y}{4\pi}+\lambda v \frac{1-\gamma}{\gamma}, \\
A_{54} &=&-\frac{B_x h_z}{4\pi}+\lambda w \frac{1-\gamma}{\gamma}, \\
A_{56} &=&-\frac{v B_x}{4\pi}+\lambda q_y,\ A_{57} =-\frac{w
B_x}{4\pi}+\lambda q_z, \\
A_{61} &=&\frac{v B_x}{\rho}-\lambda h_y,\
A_{71}=\frac{w B_x}{\rho}-\lambda h_z \end{eqnarray*}

The following notions are adopted in the above relations:
$\rho=\sqrt{\rho_1 \rho_2}$, $u=\overline{\sqrt{\rho} u}/\overline{\sqrt{\rho}}$,
$v=\overline{\sqrt{\rho} v}/\overline{\sqrt{\rho}}$, $w=\overline{\sqrt{\rho}
w}/\overline{\sqrt{\rho}}$, ${\cal H}=\overline{\sqrt{\rho}
H}/\overline{\sqrt{\rho}}$,
$h_y=\overline{B_y/\sqrt{\rho}}/\overline{\sqrt{\rho}}$,
$h_z=\overline{B_z/\sqrt{\rho}}/\overline{\sqrt{\rho}}$, where $\overline{f}$ means
arithmetic averaging. Besides,

\begin{eqnarray*}
\medmuskip=0mu
q &=&
\frac{2-\gamma}{4 \pi \gamma} \Bigl(\overline{Y}^2+\overline{Z}^2+
\frac{\theta_1}{4} (\Delta Y)^2 + \frac{\theta_2}{4} (\Delta Z)^2\\
&+&\frac{\eta_1}{4 \overline{R}}
\overline{Y} \Delta Y \Delta R +\frac{\eta_2}{4 \overline{R}} \overline{Z}
\Delta Z \Delta R \Bigr), \\
q_y &=& \frac{2-\gamma}{4\pi\gamma}\Bigl(
\overline{Y}\,\overline{R}+\frac{1-\theta_1}{4} \Delta Y \Delta R +
\frac{1-\eta_1}{4\overline{R}}(\Delta R)^2 \overline{Y}\Bigr), \\
q_z &=&
\frac{2-\gamma}{4\pi\gamma}\Bigl(
\overline{Z}\,\overline{R}+\frac{1-\theta_2}{4} \Delta Z \Delta R +
\frac{1-\eta_2}{4\overline{R}}(\Delta R)^2 \overline{Z}\Bigr),
\end{eqnarray*}

\noindent
where $\theta_1$, $\theta_2$, $\eta_1$, and $\eta_2$ are arbitrary
parameters. Their origin is caused by the presence in the expressions for
$\Delta {\bf F}$ of the terms containing the factors $\Delta R \Delta Y$
and $\Delta R \Delta Z$ which can be attributed both to the terms
proportional to $\Delta R$ and $\Delta Y$ or $\Delta Z$. This results in an
additional parametrization of the entries of the matrices $A_F$ and $A_U$.
It is not difficult to find that

\begin{eqnarray*}
\det A_\lambda &=&
\frac{2K}{\gamma}\Bigl(K^2-\frac{B_x^2}{4\pi\rho}\Bigr)\Bigl\{(K^2-
c^2-\alpha) \\
&\times& \Bigl(K^2-\frac{B_x^2}{4\pi\rho}\Bigr)
-K^2\bigl[\frac{\rho}{4\pi} (h_y^2+h_z^2) +\beta\bigr]\Bigr\},
\end{eqnarray*}

\noindent where

\begin{eqnarray*}
K &=& u-\lambda ,\ c^2=(\gamma -1) \Bigl[ I-\frac{u^2+v^2+w^2}{2} \\
  &-& \frac{B_x^2}{4 \pi \rho}-\frac{(h_y^2+h_z^2)\rho}{4\pi} \Bigr], \\
\alpha &=& \gamma\,(\delta-\delta_y h_y-\delta_z h_z)/2 =\gamma (q - q_y
h_y- q_z h_z), \\
\beta &=& \gamma (h_y \delta_y + h_z \delta_z),\ \delta=q -
\frac{2-\gamma}{4\pi\gamma}\left(h_y^2+h_z^2\right) \rho,\\
\delta_y &=& q_y -\frac{2-\gamma}{4\pi\gamma} h_y \rho,\
\delta_z=q_z -\frac{2-\gamma}{4\pi\gamma} h_z \rho
\end{eqnarray*}

The equation for $\alpha$ can be rewritten in form

\begin{eqnarray*}
\alpha &=& \frac{2-\gamma}{32\pi}\Bigl[\theta_1 (\Delta Y)^2 +
(\theta_1+\eta_1-1)\frac{\overline{Y} \Delta Y \Delta R}{\overline{R}} \\
&+& (\eta_1 -1)\Bigl(\frac{\overline{Y} \Delta
R}{\overline{R}}\Bigr)^2+\theta_2 (\Delta Z)^2 \\
&+&(\theta_2 +
\eta_2-1)\frac{\overline{Z} \Delta Z \Delta R }{\overline{R}} +
(\eta_2-1)\Bigl(\frac{\overline{Z} \Delta R}{\overline{R}}\Bigr)^2\Bigr]
\end{eqnarray*}

The eigenvalues of $A$ are equal to $u$, $u \pm b$, where
$b=|B_x|/\sqrt{4\pi\rho}$, and to the four roots of the biquadratic equation

\begin{equation}
K^4-2pK^2+Q=0,
\end{equation}

\noindent
where

\begin{eqnarray*}
2p &=& c^2+\alpha+b^2+(h_y^2+h_z^2)\rho/4\pi + \beta, \\
Q &=& (c^2+\alpha)b^2
\end{eqnarray*}

If $c^2+\alpha\ge 0$ and $(h_y^2+h_z^2)\rho/4\pi +\beta \ge 0$, the roots
of this equation are real and the diagonal matrix composed of the
eigenvalues acquires the form

\[
\Lambda = \diag ||u+a_f,\ u+b,\ u+a_s,\ u,\ u-a_s,\ u-b,\ u-a_f||
\]

The roots $a_f$ and $a_s$ are the largest and  the least root of Eq.~(39) (
the fast and the slow magnetosonic waves) and $b$ corresponds to the
Alfv\'enic waves. The remaining eigenvalue corresponds to the entropy waves.

The peculiarity of our approach lies in the strict ordering of the
eigenvalues. This provides the absence of their additional nonphysical
degeneration which is not inherent in $J$. Note that the choice of other
parameter vectors $\bf s$ can break this property. In particular, such a
degeneration appears if $q_y$ and $q_z$ are not proportional to
$\overline{Y}$ and $\overline{Z}$, respectively. This leads to the most
simple admissible choice of $\theta$: $\theta_1=\theta_2=1$.
In the MHD case, in contrast to pure gas dynamics, it is not possible to
construct the matrix $A$ depending on a single average
vector.

Let us calculate $\Omega_R$ and $\Omega_L$. It is convenient to introduce
the matrix $\Omega_r$ instead, for which $\Omega_R=A_U \Omega_r$. This
matrix consists of seven columns $\bf r$. For the eigenvalues
$\lambda=u+sa$, where $s=\pm 1$ and $a=a_f$, $a_s$, or $0$, the
corresponding vector-columns ${\bf r}={\bf r}(s,a)$ are the following:

\[
{\bf r}=\left(1,\,u+2sa,\,v-sh_yM,\,w-sh_zM,\,r_5,\,h_yN,\,h_zN\right)^T
\]

\noindent
where
\begin{eqnarray*}
r_5 &=&-{\cal H}+u^2+v^2+w^2+2sau-(vh_y+wh_z)sM \\
    &+& \bigl[2a^2-q-(q_yh_y+q_zh_z)N\bigr]\gamma/(\gamma -1), \\
M &=& M(a)=\frac{aB_x}{2\pi(a^2-b^2)},\ N=N(a)=\frac{a^2+b^2}{a^2-b^2}
\end{eqnarray*}

For the eigenvalues $\lambda=u+sb$ ($s=\pm 1$) the corresponding vector
acquires the form

\[
{\bf r}=\left(
\begin{array}{c}
0 \\
0 \\
h_z^{\ast} \\
-h_y^{\ast} \\
vh_z^{\ast}-wh_y^{\ast} \\
-sh_z^{\ast}\sqrt{4\pi/\rho}\sgn B_x \\
sh_y^{\ast}\sqrt{4\pi/\rho}\sgn B_x
\end{array}
\right)
\]

\noindent
where $h_y^{\ast}=h_y/|{\bf h}|$ and $h_z^{\ast}=h_z/|{\bf h}|$.

When using the above formulas for $|{\bf h}| \to 0$, the indeterminacies of
the type $0/0$ must be resolved. This can be done, e.~g., by the
substitution $h_y=|{\bf h}|\sin{\varphi}$ and $h_z=|{\bf h}|\cos{\varphi}$.

The matrix $\Omega_L$ can be found similarly by introducing $\Omega_l$ such
that $\Omega_L=D^{-1}\Omega_l$, where $D$ is a diagonal matrix specified by
the equality $A_U\Omega_R D^{-1}\Omega_l=I$. It consists of seven rows.
For the eigenvalues $\lambda=u+sa$, where $s=\pm 1$ and $a=a_f$, $a_s$, or
$0$, the corresponding vector-rows ${\bf l}={\bf l}(s,a)$ are the following:
\begin{eqnarray*}
l_1 &=& \frac{c^2-a^2+sau}{\gamma-1}-\frac{u^2+v^2+w^2}{2} \\
&-& 2s\pi (v r_y+w r_z) M + (h_yr_y+h_zr_z) L, \\
l_2 &=& u-\frac{sa}{\gamma -1},\ l_3 = v+2\pi Msr_y, \\
l_4 &=& w+2\pi Msr_z,\ l_5=-1, \\
l_6 &=& -r_yL+\frac{\rho h_y}{4\pi}, \ l_7=-r_zL+\frac{\rho h_z}{4\pi} \\
r_y &=& \frac{h_y}{4\pi}+\frac{\gamma q_y}{\rho (\gamma -1)},\\
r_z &=& \frac{h_z}{4\pi}+\frac{\gamma q_z}{\rho (\gamma -1)},\ L(a) =
\frac{[1+N(a)]\rho}{2}
\end{eqnarray*}

For $\lambda=u+sb$ ($s=\pm 1$) we obtain
\begin{eqnarray*}
l_1 &=&wh_y^{\ast}-vh_z^{\ast},\ l_2=l_5=0,\ l_3=h_z^{\ast},\ l_4
=-h_y^{\ast},\\
l_6 &=&-sh_z^{\ast}\sqrt{\rho/4\pi}\sgn B_x,\
l_7=sh_y^{\ast}\sqrt{\rho/4\pi}\sgn B_x
\end{eqnarray*}

The matrix $D$ has the form

\[
D=\diag ||d(a_f),\,2,\,d(a_s),\,-d(0),\,d(a_s),\,2,\,d(a_f)||,
\]

\noindent
where
\begin{eqnarray*}
d(a) &=&
-\frac{2}{\gamma
-1}\Bigl\{a^2+c^2+\alpha \\
&+& [(h_y^2+h_z^2)\rho/4\pi+\beta](1+N)N/2 \Bigr\}
\end{eqnarray*}

In practice, if $B_x \to 0$ or $|{\bf h}| \to 0$, the indeterminacy of the
type $0/0$ must be resolved in the above relations for $M=M(a)$ and
$N=N(a)$ at $a=a_s$. This can be done by using the biquadratic equation
for the roots.  It is not difficult to find that
\begin{eqnarray*}
N(a_s) &=& -\frac{(a_f^2-b^2)(c^2+\alpha +2p+b^2)}{\varepsilon |{\bf
h}|^2(a_f^2+b^2)}, \\
M(a_s) &=& -\frac{(a_f^2-b^2)\sgn B_x}{\varepsilon a_f |{\bf
h}|^2} \sqrt{\frac{(c^2+\alpha)\rho}{\pi}},
\end{eqnarray*}

\noindent
where

\[
\varepsilon=\frac{1}{4\pi} \Bigl\{\rho+\frac{(2-\gamma)(\Delta R)^2 }{4}
\bigl[(2-\eta_1)(h_y^{\ast})^2+(2-\eta_2)(h_z^{\ast})^2\bigr]
\Bigr\}
\]

To resolve the indeterminacy, eigenvectors ${\bf r}(\pm 1,~a_s)$ and
${\bf l}(\pm 1,~a_s)$ are multiplied by $|{\bf h}|$, and the corresponding
$d(a_s)$ by $|{\bf h}|^2$. Then, the substitution is made similar to that
used for the regularization of the Alfv\'enic eigenvectors.

No nonphysical degeneration occurs for $\gamma=2$, since in this case
$\alpha\equiv 0$ and $\beta\equiv 0$ and the characteristic equation has
only real roots. If we choose $\theta_1=\theta_2=1$ and $\eta_1=\eta_2=2$
for an arbitrary $\gamma$, see~\cite{kn:Hanawa}, one can easily show that
$\alpha \geq 0$ and $\beta \equiv 0$, thus giving only real roots of the
characteristic equation for any admissible right and left values.

Another choice is $\theta_1=\theta_2=1$ and $\eta_1=\eta_2=0$. In this case
$\beta \geq 0$, whereas $\alpha \geq 0$ only for the right and the left
values connected via the Hugoniot-type conditions. The possible
degeneration of the formula can be avoided by the regularization. It is
worth mentioning that the above manipulations related to the matrix
$A_\lambda$ were performed for arbitrary~$q$, $q_y$, and~$q_z$. This allows
us, by substituting $q^{\ast}$ for $q$, where

\[
q^{\ast}=|q-h_yq_y-h_zq_z|+h_yq_y+h_zq_z,
\]

\noindent
and leaving $q_y$ and $q_z$ unchanged, to conserve the properties
of $J$ universally for any $0\leq\eta_1 =\eta_2\leq 2$. For small jumps
this regularization introduces the error of the second order of smallness
and does not prevent its application for constructing the solutions of the
second order of accuracy.  It does not distort the solution for the MHD
jumps, since in this case $\alpha \geq 0$.  We also have $\beta\geq0$
for the above choice of $\eta_1$ and $\eta_2$. Such an approach preserves
the jump relations and is intermediate between the techniques using the
exact expression for $A= \Delta {\bf F}/ \Delta {\bf U}$ \cite{kn:Brio},
\cite{kn:Hanawa}, and~\cite{kn:Roe81} and approaches using
different approximations to $A$.

The family of approximate solutions to the MHD Riemann problem presented
here generalizes the known approximate quasi-linearized and linearized
solutions of this problem and preserves the Hugoniot-type relations on the
jumps. By using proper reconstruction techniques, we can increase the
order of accuracy of obtaining the fluxes in Eq.~(38). In this case the
indices ``1'' and~``2'' must be attributed to the parameter values on the
right and on the left side of the computational cell.

Due to the complexity of formulas, this approach is much more cumbersome
than Roe's linearization method in pure gas dynamics. Recently in
\cite{kn:Dai94a}, a nonlinear approximate Riemann
problem solver is suggested in which all the waves emanating from the
initial discontinuity are treated as discontinuous jumps.  That is why, it
is applicable only for weak rarefactions. Moreover, the solver
proposed is somewhat time-consuming and sensitive to the initial
approximation for the iteration process.

Taking into account the above-mentioned remarks some simplified approaches
are welcome which should (1) satisfy TVD property and (2) be enough
economical and robust.

In \cite{kn:Barmin}, the second order of accuracy in time and space TVD
Lax--Friedrichs type scheme is suggested that gives a great
simplification of the numerical algorithm in the finite-volume formulation
comparing with the schemes which use the precise characteristic splitting of
Jacobian matrices. The results obtained by this scheme were compared
with those from \cite{kn:Brio} and \cite{kn:Dai94b} and a good agreement
was observed.  In this scheme we substitute the diagonal eigenvalue
matrix in Eq.~(38) by the diagonal matrix with the spectral radius~(the
maximum of eigenvalue magnitudes) of $A$ on its diagonal.
Using the proper parameter reconstruction to find their values on the
computational cell surfaces, we obtain the second order of accuracy.

This
scheme is less dissipative than the original Lax--Friedrichs scheme and can
be applied to calculations of discontinuous MHD flows. Being much simpler
than the scheme based on Roe's linearization method, it still gives
numerical results with the reasonable accuracy.

Another important subject of  discussion  is that certain
initial- and boundary-value problems
can be solved nonuniquely using different shocks or different combinations of
shocks, whereas physically one would expect only unique solutions. The
situation differs from that in pure gas dynamics, where all
entropy-increasing solutions are evolutionary and physically admissible.
This means that the  necessary conditions of the well-posedness for the
linearized problem of their interaction with small disturbances are
satisfied. In MHD case, on the contrary, the condition of the entropy
increase is necessary, but not sufficient.  Only slow and fast MHD shocks
turned out to be evolutionary, while intermediate (or improper slow) shocks
are to be excluded \cite{kn:Jeffrey} and \cite{kn:Kulik65}. Nonevolutionary
shocks are not simply unstable in ideal MHD. Their decay into evolutionary
jumps occurs under infinitesimal perturbation within  infinitesimal time.
On the other hand, numerical viscosity (including the presence of finite
conductivity) and numerical dispersion of a numerical scheme make such
intermediate structures existent for a certain time interval before their
destruction. It is very important to realize that this interval has
nothing to do with the real interval of existence of intermediate waves
in non-ideal plasma and is grid- and numerical scheme-dependent.
If the viscosity and/or the conductivity are substantial, such compound
structure can exist for a long time depending on the amount of viscosity.
This was admitted and used for the explanation of certain physical
phenomena in~\cite{kn:Wu}.

In \cite{kn:Brio}, the ideal MHD Riemann problem
was solved with  initial data consisting of two constant states lying to
the right and to the left of the centerline of the computational domain.
Being solved as a strictly coplanar problem, it included a nonevolutionary
compound shock. Such shocks
must decay and are not realizable in physical problems.
The peculiarity of MHD is that there exist discontinuities that are
nonevolutionary only with respect to Alfv\'enic (rotational)
disturbances.
That is why, if a strictly coplanar problem is considered (velocity and
magnetic field vectors lie in the same plane and the system of MHD
equations includes only two vector components) the construction of the
solution is possible both with evolutionary and nonevolutionary shock
waves. The solution in this case is non\-unique. If the full set of
three-dimensional MHD equations is solved and a small tangential
disturbance is added to the magnetic field vector, a rotational jump
splits from the compound wave and it degrades into the slow shock. This
means that the compound wave is unstable against tangential disturbances
and is nonevolutionary in three dimensions (see \cite{kn:Barmin}).
That means that one must be very careful reducing the dimension of the system
(say, in axisymmetric problems) to avoid the origin of nonadmissible
solutions. The necessity of three-dimensional consideration of MHD Riemann
problems has been  admitted recently in \cite{kn:Dai94a}.
One must take into account this feature of MHD
equations in the construction of numerical algorithms. In other words, if we
are going to solve an axisymmetric problem, in general, a full set of
three-dimensional equations must be used.

\subsection{Shock-fitting calculations}

In \cite{kn:BarZa} the interaction between the solar wind and
the magnetized plasma component of the local interstellar medium were
calculated by solving the axisymmetric MHD equations using the shock-fitting
approach. The solar wind was
assumed nonmagnetized. The magnetic field of the interstellar medium was
assumed parallel to its velocity vector. The system of MHD equations in
the quasi-linear form   was solved by the iteration method. Each
iteration consisted of the two steps. At the first step, the gasdynamic
part of the system was solved assuming magnetic field known from the
previous time step. At the second iteration step, the Maxwell equations
were solved assuming velocities known. At the first step of each iteration
magnetic field was assumed zero.  The gasdynamic part was solved by the
time-stabilization method \cite{kn:Zaitsev} (see Section~3). For the
known distribution of gasdynamic parameters the magnetic field in the
considered case can be determined analytically as
\[
{\bf B}= B_\infty {\rho {\bf v} \over \rho_\infty V_\infty}
\]

These equation follows from the induction equations and the MHD jump
conservation relations \cite{kn:Landau} for ${\bf B}\parallel {\bf v}$.

The following solar wind and LISM parameters were chosen:
\begin{eqnarray*}
& &n_e= 7\, {\rm cm}^{-3},\ V_e=450\,{\rm km/s},\ M_e=10, \\
& &n_\infty=0.07\, {\rm cm}^{-3},\ V_\infty=25\,{\rm km/s},\ M_\infty=2
\end{eqnarray*}

The dimensionless value of magnetic field was specified via parameter
$\alpha=2\sqrt\pi/A$, where $A=V_\infty/\sqrt{B_\infty/4\pi\rho}$ is the
Alfv\'en number.

The geometrical pattern of the flow is shown in Fig.~19.

\begin{figure}[t]
\centerline{\psfig{figure=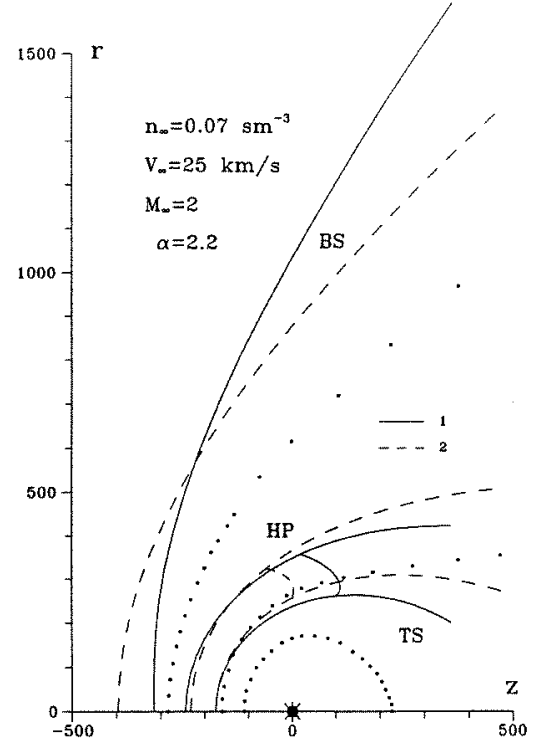,height=10.0cm}}
\vspace{-6pt}
\caption{(1)
Geometrical pattern of the interface for $\alpha=2.2$ [13].
Positions of the bow shock (BS), the termination shock (TS), heliopause
(HP), and sonic line in the solar wind. (2) The same lines for $\alpha=0$.
The discontinuities calculated in [11] in the presence of the
neutral hydrogen are shown by dotted lines}

\vspace{-12pt}
\end{figure}

It is quite clear from this picture that both the LISM magnetic field and
the charge-exchange processes play an important role in the
interaction. One can notice that, although a shock-fitting approach
perfectly well works in the forward part of the interaction, its
limitations reveal themselves in the tail part of the flow. The
applicability of the described approach is also limited to axisymmetric
problems.

\subsection{Shock-capturing calculations}

All mentioned in the previous subsection necessitates the application of
shock-capturing methods for modeling of the SW--LISM interaction.
In~\cite{kn:Mats93} this problem was solved in the closed region
surrounding an ejecting star on the basis of the flux-splitting
method~\cite{kn:Jyo}. The choice of parameters was far from
those adopted nowadays for the considered problem. Although, as a whole, a
physically consistent results were obtained, the resolution of
discontinuities was rather poor and some of the obtained data were
misinterpreted and disputed in~\cite{kn:BarZa}.

In \cite{kn:Po97}  the solution of the axisymmetric problem (the
LISM magnetic field strength vector is assumed to be parallel to its
velocity vector) is presented for realistic SW and LISM parameters. The
numerical method applied was developed by
Pogorelov \cite{kn:Barmin} and is the high-resolution
second-order of accuracy version of the Lax--Friedrichs scheme. This
method gives a drastic simplification of the numerical algorithm
comparing with the methods based on the exact characteristic splitting
of the Jacobian matrices in the MHD equations.  Simple but
very effective numerical boundary conditions at the far-field were
suggested (see Section~2),
which allowed one to avoid the influence of spurious reflections from the
subsonic outer boundary.  An effective
procedure of satisfying the condition of source-free magnetic
fields (divergence-free condition) was used, which is related to the
approach \cite{kn:Powell94}.

The formulation of the problem was given is Section~2. Following
\cite{kn:Barmin}, the numerical flux at the radial cell interface (27) was
calculated by the formula
\begin{eqnarray}
& &{\bar {\bf E}}_{l+1/2, n} = \frac {1}{2} \left[{\bar {\bf E}}\left({\bf
  U}^R_{l+1/2,n} \right) + {\bar {\bf E}}\left({\bf U}^L_{l+1/2,n}\right)
  +{\bf \Phi}_{l+1/2,n}\right]\, , \nonumber \\
& &  \\
& &{\bf \Phi}_{l+1/2,n} = -
  {\bf {\hat R}}_{l+1/2,n} \left({\bf U}^R_{l+1/2,n} - {\bf U}^L_{l+1/2,n}
   \right) \nonumber
\end{eqnarray}

Here ${\bf {\hat R}}_{l+1/2,n}$ is the diagonal matrix with the same
elements on its diagonal equal to the spectral radius r (the maximum of
eigenvalue magnitudes) of the Jacobian matrix
${\partial {\bar {\bf E}} \over \partial {\bf U}}$:
\begin{eqnarray*}
& &r=|U|+a_f,\quad a_f^2={1 \over 2}
\left( (a^{*})^2+ \sqrt{(a^{*})^4-4a^2 b_R^2} \right)\, , \\
& &b_R=B_R/(4\pi \rho)^{1/2},\quad B^2=B_R^2+B_\theta^2\, , \\
& &(a^{*})^2=(\gamma p + {B^2 / 4 \pi})/\rho, \quad a^2=\gamma p/\rho\, ,
\end{eqnarray*}

\noindent
where $U$ is the radial velocity component and $B_R$ and $B_\theta$ are the
radial and the angular component of the magnetic field strength vector.
Similar formulas can be written for the fluxes in the angular direction.

Having the second order of accuracy, the proposed
scheme is much less dissipative than the original Lax--Friedrichs method
and provides incomparably better shock resolution.
Calculations were performed  in the polar computational
region with $R_{\rm min}=24$ and $R_{\rm max} = 1200$. The mesh was $R
\times \theta = 99 \times 116$.

Mathematically, if we choose the initial distribution of parameters with
$\div {\bf B} = 0$, the Maxwell equations will preserve this value in the
steady solution. In fact, for problems solved numerically the regions
of $\div {\bf B} \neq 0$  can be accumulated, especially
in the vicinity of discontinuities, see \cite{kn:Brackbill80}. It is
quite clear that the application of the one-dimensional Riemann problem
solvers, implying that the component $B_n$ of the magnetic field vector
normal to the boundary is constant, contradicts to  the condition
$\oint B_n d \sigma = 0$ over the whole computational cell.
In \cite{kn:Powell94} an approximate Riemann problem solver for MHD equations
was proposed  on the basis of the modified system~(35) which is
conservative only in a steady state.  Using the Lax--Friedrichs-type
scheme, we do not need such modification, since we do not
solve  the Riemann problem to find the fluxes at the cell surfaces. The
correction in our case can be made by adding to the source term of
Eq. (2) the value proportional to $\div {\bf B}$:
$$
{\bf H}' =\div {\bf B}\,\left[0,\,{B_x\over 4\pi},\,
 {B_z\over 4\pi},\,{{\bf v}\cdot {\bf B}\over 4\pi},\,u,\,w\right]^{\rm T}
$$

This term acts to annihilate the error accumulated if system is
solved in the conservation-law form (see \cite{kn:Brackbill80}). It is
worth noting that this approach lies in the framework of Powell's
procedure  and the correction term is not small only in the regions of
comparably large errors  in $\div {\bf B}$. This term, in fact, is equal
by the value and opposite by the sign to the appropriate terms
proportional to $\div {\bf B}$ appearing by differentiating $\bf E$ and
{\bf G} in Eq.~(1). The divergence of
magnetic field strength vector can be approximated over the finite volumes as
\begin{eqnarray*}
& &\div {\bf B} = [R^2_{l+1/2}(B_n)_{l+1/2,n} +
R^2_{l-1/2}(B_n)_{l-1/2,n}]/R^2_l \Delta R + {} \\
& & [(B_n)_{l,n+1/2} \sin \theta_{n+1/2} + (B_n)_{l,n-1/2} \sin
\theta_{n-1/2}]/R_l \sin \theta_n \Delta \theta
\end{eqnarray*}

This simple procedure gives a powerful tool for the realization of the
divergence-free condition and necessitates only a slight modification of
the existing codes for solving MHD equations. It seems limited, however,
only to steady-state calculations.

Numerical results were obtained for the same flow parameters as in the
previous subsection and for the three
different  values of the Alfv\'en number: $A = 10,\, 2,\, \sqrt{2}$. The
first value corresponds to the very small magnetic field (see Fig.~20)
and  can be used for comparison with purely gasdynamic results in which the
Roe-type Riemann problem
solver  was used to obtain the steady state solution, see Subsection
3.2.  In Fig.~20,
the pressure (below the symmetry axis) and the density logarithm isolines
are presented ($A=10$).  All features of the flow
pattern  (see Fig.~2) are sharply resolved, while the low time-consumption
and  the simplicity of the algorithm are quite clear.
In  Fig.~21 the same distributions are shown for the case
$A=2$.

\begin{figure}[t]
\centerline{\psfig{figure=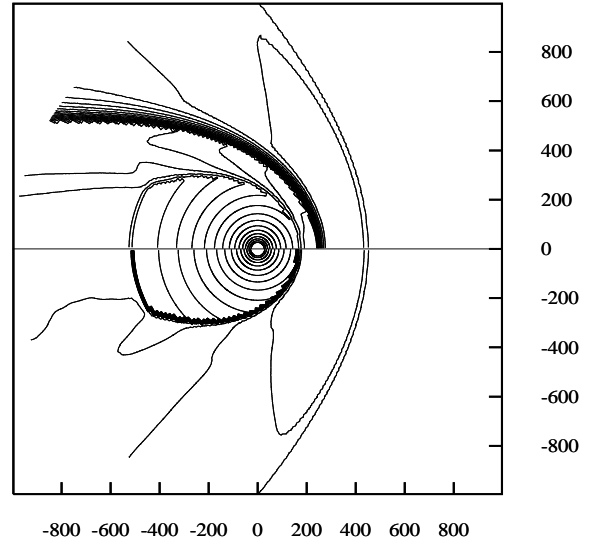,height=7.15cm,width=7.7cm}}
\vspace{-6pt}
\caption{
Pressure (below the symmetry axis) and density logarithm
isolines,\ $A=10$
}
\vspace{-12pt}
\end{figure}

In
Figs.~22--23, the case with $A = \sqrt{2}$ is presented, which
corresponds to
$p_{\rm magn}/p_{\rm thermal} \approx  1. 67$ at infinity, that is,
$B_\infty \approx  2.5$ (dimensional value $\sim 2.3\times 10^{-6}$
Gauss). In Fig.~23, the streamlines (lower half) and the magnetic
field lines are shown.

\begin{figure}[t]
\centerline{\psfig{figure=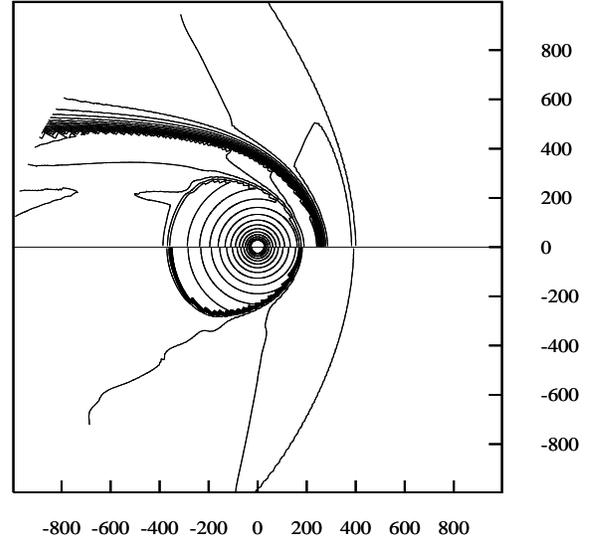,height=7.15cm,width=7.7cm}}
\vspace{-6pt}
\caption{
Pressure (below the symmetry axis) and density logarithm
isolines,\ $A=2$
}
\vspace{-12pt}
\end{figure}

\begin{figure}[t]
\centerline{\psfig{figure=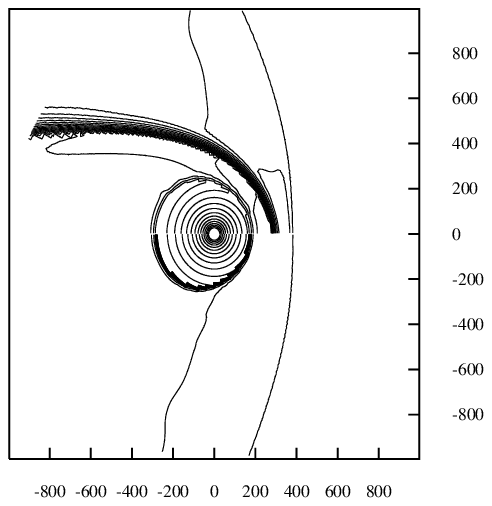,height=7.15cm,width=7.7cm}}
\vspace{-6pt}
\caption{
Pressure (below the symmetry axis) and density logarithm
isolines,\ $A=1.414$
}
\vspace{-12pt}
\end{figure}

\begin{figure}[t]
\centerline{\psfig{figure=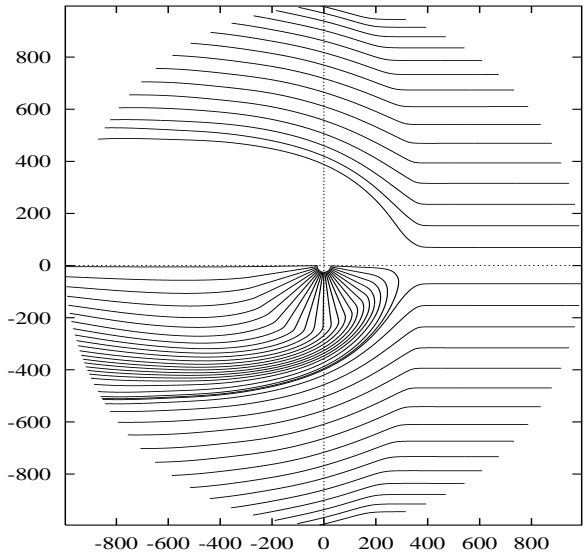,height=7.26cm,width=7.7cm}}
\vspace{-6pt}
\caption{
Streamlines (below the symmetry axis) and magnetic field lines, \
$A=1.414$.
}
\vspace{-12pt}
\end{figure}

\begin{figure}[t]
\centerline{\psfig{figure=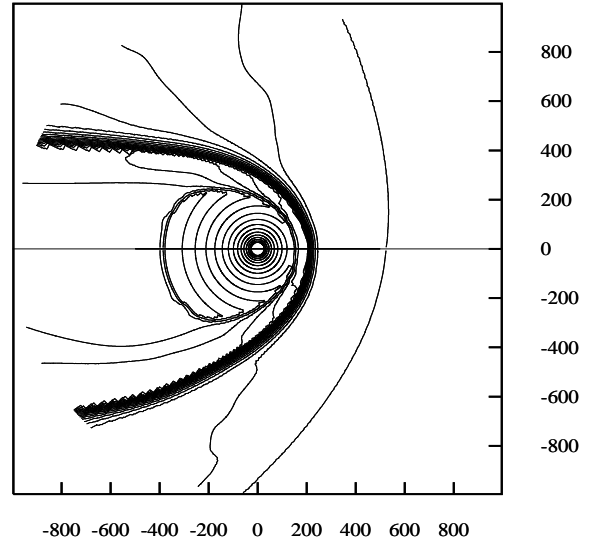,height=7.15cm,width=7.7cm}}
\vspace{-6pt}
\caption{
Density logarithm isolines in the symmetry plane
}
\vspace{-12pt}
\end{figure}

Near the
symmetry  axis the heliocentric distance of the heliopause increases due
to  the magnetic field tension, whereas it decreases at the side parts
due to the magnetic pressure. On the other hand, although the magnetic
field  is absent in the region inside the heliopause, its action is
revealed by the increased value of the total pressure at infinity. This
leads to substantial decrease of the termination shock stand-off distance
in  the downstream region.
We can see the absence of the Mach disk structure in the backward
direction. This means that the velocity along the termination shock
becomes  subsonic. The same occurs if the charge-exchange is taken into
account~\cite{kn:Bar93}. The bow shock  stand-off distance along the
symmetry axis becomes smaller than in the absence of magnetic field.

As the magnetic field increases, the effective Mach number, generally
speaking, diminishes.
The similar effect is produced by the
presence of neutral hydrogen atoms and, if both of these factors exceed
a  definite value, the bow shock can disappear. In this case system (2)
becomes elliptic and other methods must be applied for its solution.
It is worth noting, however, that the speed of
propagation of magnetosonic waves depends on the direction with respect
to the magnetic field vector. That means that the effective Mach number
varies along the bow shock. As will be shown later, this results in a highly
asymmetric shape of the bow shock if magnetic field is not parallel to
the velocity vector.

If the magnetic field strength at infinity  is equal to its probable
upper limit of $3 \times 10^{-6}$ Gauss,
the LISM flow becomes subsonic, even if the charge-exchange processes are
neglected. This can be seen from the simple estimate. We can calculate
the LISM effective Mach number as $M^2_{\rm eff}=\rho_\infty V^2_\infty/
(\gamma p_\infty+B^2_\infty/4\pi)$ (this holds, e. g., at the symmetry
axis for ${\bf V}_\infty\perp {\bf B}_\infty$  .
Thus,$1 / M^2_{\rm eff}= 1 / M^2_\infty +1/ A^2$
and for $M^2_\infty=4$ the effective Mach number remains larger than
unity only for $A>1.15$.
Thus, both effects are of great importance for the
interpretation  of data obtained in the space experiment.
The performance of the algorithm for the realization of the magnetic
field source-free condition can be seen from Fig.~23. The
magnetic field lines remain parallel to the streamlines.  One can see
that no magnetic field penetrates into the heliosphere.  This result can
not be achieved without a special treatment of the divergence-free
condition.

The result of the magnetic field influence on the whole structure
of the flow  is not sufficiently investigated.
It is clear that the flow pattern for ${\bf B}_\infty \not\parallel {\bf
V}_\infty$  is three-dimensional. MHD modeling of the
heliopause shape on the basis of the Newtonian approximation was
performed in \cite{kn:Fahr88}. The results of the three-dimensional modeling
of the solar wind interaction with the magnetized interstellar medium were
presented in~\cite{kn:PoMats}.

\begin{figure}[t]
\centerline{\psfig{figure=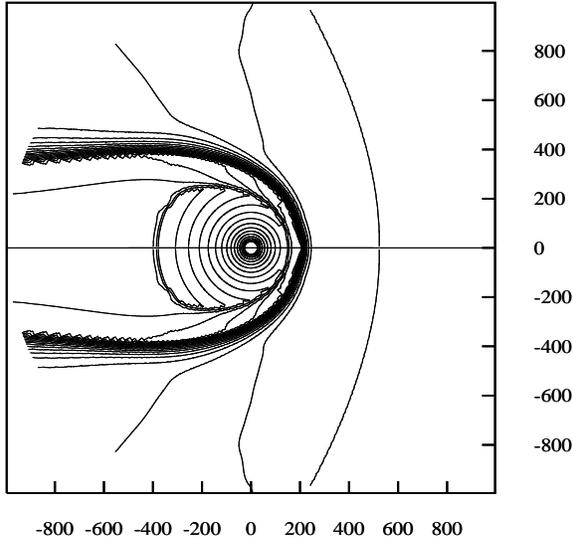,height=7.15cm,width=7.7cm}}
\vspace{-6pt}
\caption{
Density logarithm isolines in the plane $\varphi=90^{\circ}-270^{\circ}$
}
\vspace{-12pt}
\end{figure}

Calculations were performed  in the spherical computational
region with $R_{\rm min}=24$ and $R_{\rm max} = 1200$. The mesh is $R
\times \theta \times \varphi = 99 \times 116 \times 21$.
All parameters are the same as in the previously described axisymmetric
calculation and $A=2$.
In Figs.~24--25 the logarithm density isolines are shown for the case of
the angle between ${\bf v}_\infty$ and ${\bf B}_\infty$ equal to $\alpha=45^\circ$
in the cross-sections $\varphi=0\hbox{--}180^\circ$  and
$90^{\circ}\hbox{--}270^\circ$, respectively. The LISM influences in this
case the shape of the bow shock as well as the location of the stagnation
point at the heliopause surface in a way consistent with the simplified study
\cite{kn:Fahr88}. In addition, the results show the existence
of magnetosheath current layers providing proper rotation of the magnetic
field with respect to the velocity vector from $45^\circ$ to 0 or
$180^\circ$ at the heliopause surface. The bow shock wave stand-off distance
is larger in the regions with a larger angle between the magnetic field and
the shock normal. The contact surface is substantially contracted by the
magnetic pressure in the $xy$-plane (Fig.~25) rather than in the symmetry
$xz$-plane (Fig.~24). As was mentioned earlier, the size of the zone between
the bow shock and the heliopause is very important in view of the
charge-exchange processes in this zone.

\section{ Conclusions }

In this work we presented a review of the application of numerical
methods to modeling of the stellar wind interaction with the interstellar
medium. This is only one among various domains of their application
to space simulation problems. The environment of distant stars is not so
well investigated as the solar system. That is a particular reason of an
extremely intensive study of the solar wind and the local interstellar
medium interaction. The problem is rather complicated even in gasdynamic and
MHD formulations, since the flow pattern contains a number of intersecting
discontinuities and is substantially three-dimensional. The study of this
and other astrophysical and industrial problems has recently summoned the
extension of high-resolution numerical methods to magnetohydrodynamic flows.
The peculiarity of the problem  considered in this paper is that a continuum
approach is
applicable only to the plasma component of the both flows. Trajectories of
neutral particles must be calculated using a direct Monte-Carlo
simulation. Although several authors (see brief discussion in
\cite{kn:Williams}) argue that the
application of a fluid approximation neutrals gives results close to those
obtained on the basis of the self-consistent Euler--Boltzmann formulation,
there is still a necessity to realize this possibility and to perform a
critical comparison of numerical data for the same set of defining
parameters.

The influence of the cosmic rays on the interaction is rather well known,
but their inclusion into available numerical algorithms is at the initial
stage.

We examined in this review mainly numerical aspects of the problem. This
work by no means can be considered as an exhaustive description of
physical phenomena which take place in the interaction region. Discussing the
subject, we inevitably had our preferences  and paid them more attention.
The list of references can be extended by hundreds of publications on the
considered problem, but we hope that even those mentioned in this review
give an opportunity to realize the main processes defining the problem and
approaches to their numerical modeling.

\ack
The authors are grateful to the Japanese Society of Computational Fluid
Dynamics and, personally, to Prof. Koichi Oshima who invited them to write
this review. The work of T.M. was supported by the Grant-in-Aid for
scientific research C-08640375 of the Japanese Ministry of Education,
Science, Sports, and Culture. N.P. was supported, in part, by the Russian
Foundation for Basic Research  Grant 95-01-00835.

Special thanks to K. Okuda for the help in the preparation of the graphic
material.
\vfil\eject

\bibliography{cfdlat}
\end{document}